\documentclass[twocolumn,times,twocolappendix]{aastex63}

\newcommand{\rev}[1]{#1}
\newcommand{\revtwo}[1]{#1}

\received{2020 April 23}
\revised{2020 July 1}
\accepted{2020 July 3}
\submitjournal{ApJL}

\shorttitle{A Directly Imaged Multiplanet System around TYC~8998-760-1}
\shortauthors{Bohn et al.}
\graphicspath{{./}{figures/}}

\begin{document}

\title{Two Directly Imaged, Wide-orbit Giant Planets around the Young, Solar Analog TYC~8998-760-1\footnote{
Based on observations collected at the European Organisation for Astronomical Research in the Southern Hemisphere under ESO programs 099.C-0698(A), 0101.C-0341(A), 2103.C-5012(B), and 0104.C-0265(A).
}
}

\correspondingauthor{Alexander~J.~Bohn}
\email{bohn@strw.leidenuniv.nl}

\author[0000-0003-1401-9952]{Alexander~J.~Bohn}
\affiliation{Leiden Observatory, Leiden University, PO Box 9513, 2300 RA Leiden, The Netherlands}

\author[0000-0002-7064-8270]{Matthew~A.~Kenworthy}
\affiliation{Leiden Observatory, Leiden University, PO Box 9513, 2300 RA Leiden, The Netherlands}

\author{Christian~Ginski}
\affiliation{Anton Pannekoek Institute for Astronomy, University of Amsterdam, Science Park 904, 1098XH Amsterdam, The Netherlands}
\affiliation{Leiden Observatory, Leiden University, PO Box 9513, 2300 RA Leiden, The Netherlands}

\author[0000-0003-3688-5798]{Steven~Rieder}
\affiliation{University of Exeter, Physics Department, Stocker Road, Exeter, EX4 4QL, UK}

\author[0000-0003-2008-1488]{Eric~E.~Mamajek}
\affiliation{Jet Propulsion Laboratory, California Institute of Technology, 4800 Oak Grove Drive, M/S 321-100, Pasadena CA 91109, USA}
\affiliation{Department of Physics \& Astronomy, University of Rochester, Rochester NY 14627, USA}

\author[0000-0001-6126-2467]{Tiffany~Meshkat}
\affiliation{IPAC, California Institute of Technology, M/C 100-22, 1200 East California Boulevard, Pasadena CA 91125, USA}

\author[0000-0002-7859-1504]{Mark~J.~Pecaut}
\affiliation{Rockhurst University, Department of Physics, 1100 Rockhurst Road, Kansas City MO 64110, USA}

\author[0000-0003-2911-0898]{Maddalena~Reggiani}
\affiliation{Institute of Astronomy, KU Leuven, Celestijnenlaan 200D, B-3001 Leuven, Belgium}

\author{Jozua~de~Boer}
\affiliation{Leiden Observatory, Leiden University, PO Box 9513, 2300 RA Leiden, The Netherlands}

\author[0000-0002-1368-841X]{Christoph~U.~Keller}
\affiliation{Leiden Observatory, Leiden University, PO Box 9513, 2300 RA Leiden, The Netherlands}

\author[0000-0003-1946-7009]{Frans~Snik}
\affiliation{Leiden Observatory, Leiden University, PO Box 9513, 2300 RA Leiden, The Netherlands}

\author[0000-0002-3807-3198]{John~Southworth}
\affiliation{Astrophysics Group, Keele University, Staffordshire ST5 5BG, UK}






\begin{abstract}
Even though tens of directly imaged companions have been discovered in the past decades, the number of directly confirmed multiplanet systems is still small.
Dynamical analysis of these systems imposes important constraints on formation mechanisms of these wide-orbit companions.
As part of the Young Suns Exoplanet Survey (YSES) we report the detection of a second planetary-mass companion around the 17\,Myr-old, solar-type star TYC~8998-760-1 that is located in the Lower Centaurus Crux subgroup of the Scorpius--Centaurus association.
The companion has a \rev{projected physical separation} of 320\,au and several individual photometric measurements from 1.1 to 3.8 microns constrain a companion mass of $6\pm1\,M_\mathrm{Jup}$, which is equivalent to a mass ratio of $q=0.57\pm0.10\%$ with respect to the primary.
With the previously detected $14\pm3\,M_\mathrm{Jup}$ companion that is orbiting the primary at 160\,au, TYC~8998-760-1 is the first directly imaged multiplanet system that is detected around a young, solar analog.
We show that circular orbits are stable, but that mildly eccentric orbits for either/both components ($e > 0.1$) are chaotic on Gyr timescales, implying in-situ formation or a very specific ejection by an unseen third companion.
Due to the wide separations of the companions TYC~8998-760-1 is an excellent system for spectroscopic and photometric follow-up with space-based observatories such as the James Webb Space Telescope.

\end{abstract}



\section{Introduction}
\label{sec:introduction}

Driven by the installation of extreme adaptive-optics (AO) assisted imagers such as the Gemini Planet Imager \citep[GPI;][]{macintosh2014} and the Spectro-Polarimetric High-contrast Exoplanet REsearch \citep[SPHERE;][]{Beuzit2019} instrument, the number of directly imaged extrasolar planets has been  increasing continuously over the past years.
Even though several substellar companions have been identified and characterized with these instruments \citep[e.g.][]{macintosh2015,galicher2014,chauvin2017a,keppler2018,muller2018,janson2019,mesa2019}, only two systems have been detected so far that show unambiguous evidence for the presence of more than one directly imaged companion:
one of these multiplanet systems is HR~8799 -- an approximately 30\,Myr-old star of spectral class A5 that is harboring four giant planets at orbits with semi-major axes ranging from 15\,au to 70\,au \citep[][]{marois2008,marois2010,wang2018}.
The other one is PDS~70, which is a K7-type star at an age of approximately 5.4\,Myr that is hosting at least two accreting protoplanets inside the gap of a transitional disk that is surrounding this pre-main-sequence star \citep{keppler2018,muller2018,haffert2019}.
These multiplanet systems are intriguing laboratories to study dynamical interactions and scattering events between several planetary-mass companions, which is is crucial for understanding the formation and dynamical evolution of planetary systems \citep[e.g.][]{morbidelli2018}.

To obtain a statistically significant census of wide-orbit companions to solar-type stars we launched the Young Suns Exoplanet Survey \citep[YSES;][]{bohn2020a} targeting a homogeneous sample of 70 solar-mass pre-main-sequence stars in the Lower Centaurus Crux subgroup of the Scorpius--Centaurus association \citep[Sco--Cen;][]{dezeeuw1999,pecaut2016}.
Within the scope of this survey, we already detected a self-shadowed transition disk around Wray~15-788 \citep{bohn2019} as part of a stellar binary with the debris disk host HD~98363 \citep{chen2012,moor2017,Hom2020}.
Most recent was the announcement of a $14\pm3\,M_\mathrm{Jup}$ companion that is orbiting the solar analog TYC~8998-760-1 (2MASSJ13251211–6456207) at a \rev{projected} separation of 160\,au \citep{bohn2020a}.
The primary is a $16.7\pm1.4\,$Myr-old K3IV star with a mass of $1.00\pm0.02\,M_\sun$, located at a distance of $94.6\pm0.3$\,pc \citep{bailerjones18,gaia2018}.
We refer to Table~1 of \citet{bohn2020a} for further information on the host star.

In this article we present new data on this system and report the detection of a second, farther separated, yet lower-mass companion to this young solar analog.
Sect.~\ref{sec:observations_data_reduction} outlines the observations that we acquired on TYC~8998-760-1 and how the data were reduced.
In Sect.~\ref{sec:results} we present the results of this analysis and study the properties of this gas giant companion.
Our conclusions and further prospects on characterization of this intriguing multiplanet system are presented in Sect.~\ref{sec:conclusions}.

\section{Observations and data reduction}
\label{sec:observations_data_reduction}

On the night of 2020 February 16 we acquired data on TYC~8998-760-1 with SPHERE/IRDIS \citep{dohlen2008} which was operated in dual-polarization imaging mode \citep[DPI;][]{deBoer2020,vanHolstein2020} with the instrument derotator switched off (PI: A.~Bohn).
SPHERE is mounted at the Very Large Telescope (VLT) of the European Southern Observatory (ESO) and it is supported by the SAXO extreme AO system \citep{fusco2006} to provide Strehl ratios better than 90\,\% in $H$ band.
Within the scope of this work we only used the total intensity frames of the DPI dataset that are created by adding the left and right sides of the IRDIS detector.
Furthermore, we used parts of the observations presented in \citet{bohn2020a} that were collected with NACO \citep{lenzen2003,rousset2003} and SPHERE/IRDIS in classical and dual-band imaging modes \citep{vigan2010}.
A detailed description of all observations, applied filters, and weather conditions is presented in Appendix~\ref{sec:observations_setup_and_conditions}.

The data reduction was performed as described in \citet{bohn2020a} using a custom processing pipeline based on version 0.8.1 of PynPoint \citep{stolker2019} that includes dark and flat calibration, bad pixel cleaning, and subtraction of the sky and instrument background.
A more detailed description is presented in Appendix~\ref{sec:app_data_reduction}.

\section{Results and analysis}
\label{sec:results}

\begin{figure*}
\includegraphics[width=\textwidth]{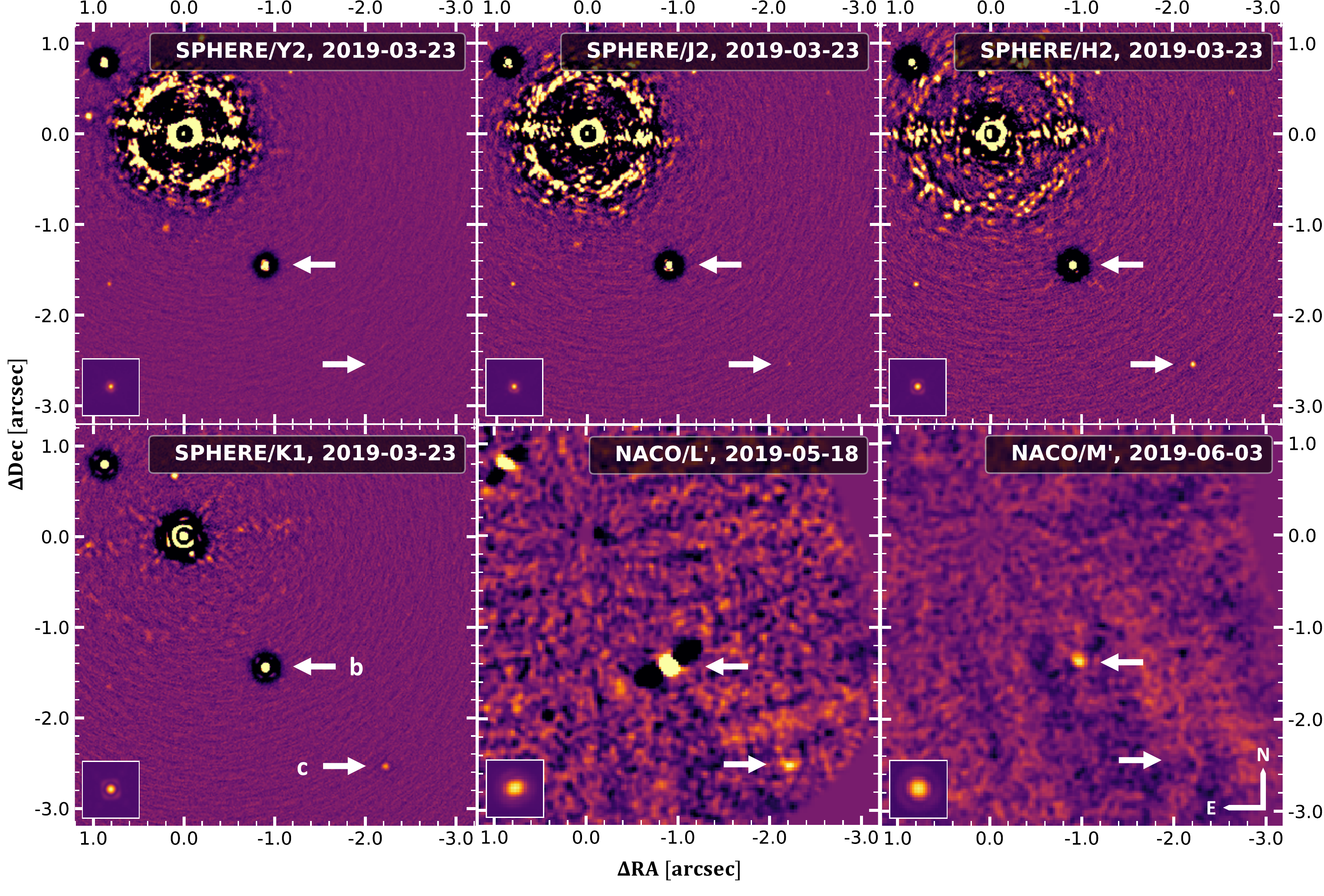}
\caption{
Two planetary-mass companions around TYC~8998-760-1.
We present the reduced data for several SPHERE and NACO filters.
The white arrows indicate the positions of the confirmed, planetary-mass companions TYC~8998-760-1~b and c as labeled in the bottom left panel.
All other objects in the field of view are background contaminants confirmed by proper motion analysis.
To highlight off-axis point sources an unsharp mask is applied to the SPHERE data and we smoothed pixel-to-pixel variations in the NACO data with a Gaussian kernel.
All images are displayed with an arbitrary logarithmic color scale.
The primary is in the upper left of each panel setting the origin of the coordinate system that represents the differential offsets in R.A. and decl.
In the lower left of each panel, we present the noncoronagraphic flux PSF as a reference for the corresponding filter.
In all frames, north points up and east is to the left.
}
\label{fig:yses_companions}
\end{figure*}

We report the detection of a second, very red companion to TYC~8998-760-1 which we will refer to as TYC~8998-760-1~c henceforth.
A compilation of both confirmed companions around this young, solar analog in several SPHERE and NACO bandpasses is presented in Fig.~\ref{fig:yses_companions}.
\rev{
TYC~8998-760-1~c was detected with a signal-to-noise ratio greater than 5 from $Y3$ to $L'$ band and we did not detect any significant flux at the expected position in the $Y2$ and $M'$ filters.
A detailed analysis of the detection significance for the individual bandpasses and nights is presented in Appendix~\ref{sec:signal_to_noise}.
}

\subsection{Astrometric analysis}
\label{subsec:astrometric_analysis}

The main confirmation of the companionship was performed by common proper motion analysis.
Because both companions are well separated from the PSF halo of the primary and no PSF subtraction was performed, we extracted the astrometry in the final images with a two-dimensional Gaussian fit.
In the $H$ band data collected on the night of 2017 July 5, we detected TYC~8998-760-1~c at a separation of $3\farcs369\pm0\farcs033$ and a position angle of $221\fdg1\pm0\fdg6$ with respect to the primary\footnote{
The uncertainties of these measurements are much larger than the usual astrometric precision of SPHERE.
This is attributed to the nonoptimal AO performance caused by poor atmospheric conditions with an average seeing of 1\farcs22 and a coherence time of 2.9\,ms, \rev{resulting in a smeared PSF and limited astrometric accuracy (see Appendix~\ref{sec:signal_to_noise}).}
}.
From the $K1$ band data -- which provides the highest signal-to-noise ratio of the companion on the night of 2019 March 23 -- we derived a separation of $3\farcs377\pm0\farcs005$ and position angle of $221\fdg2\pm0\fdg1$ east of north.
For the $H$ band data from 2020 February 16, a separation of $3\farcs380\pm0\farcs006$ and a position angle of $221\fdg3\pm0\fdg1$ were measured.
\rev{These measurements imply a projected physical separation of approximately $320$\,au at the distance of the system.}

\begin{figure}
\plotone{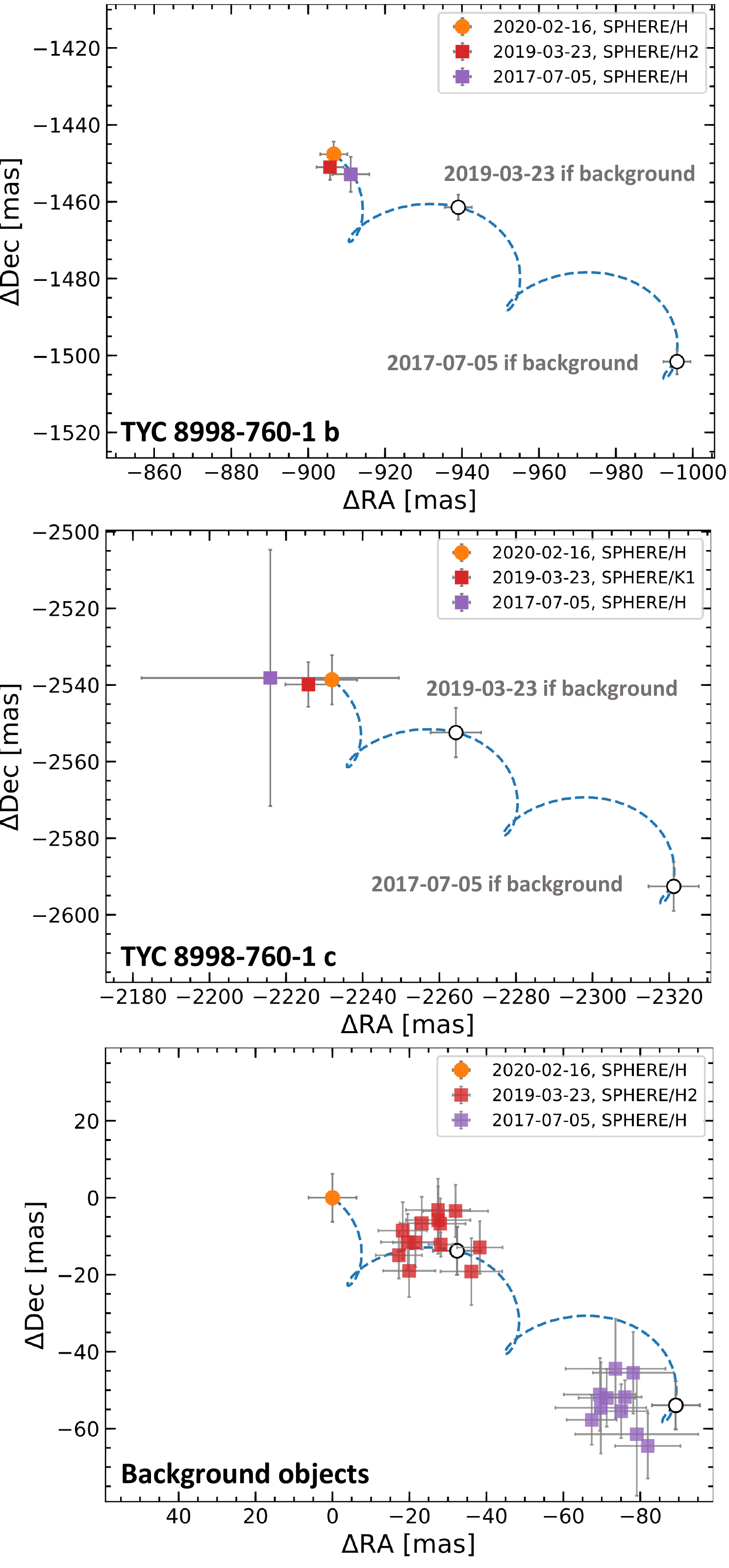}
\caption{
Multiepoch proper motion assessments of TYC~8998-760-1~b, c, and confirmed background objects.
The colored markers represent the extracted relative astrometry of objects in the SPHERE field of view.
The blue, dashed line represents the trajectory of a static background object and the white circles indicate the expected position of such an object, evaluated at the epochs indicated in the top and middle panels.
Whereas the origin of the coordinate system is located at the position of the star for the comoving companions (top and middle panel), we present the relative offsets to reference epoch 2020 February 26 for confirmed background objects (bottom panel).
The field of view sizes of the plots and the relative positions of the background trajectories are identical for all three panels, so that individual measurements of companions and background objects can be compared amongst each other.
}
\label{fig:proper_motion_diagram}
\end{figure}

This proper motion analysis is visualized in Fig.~\ref{fig:proper_motion_diagram}.
The primary has a parallax of $10.54\pm0.03$\,mas and proper motions of $\mu_\alpha=-40.90\pm0.04\,\mathrm{mas}\,\mathrm{yr}^{-1}$ and $\mu_\delta=-17.79\pm0.04\,\mathrm{mas}\,\mathrm{yr}^{-1}$ based on Gaia DR2 \citep{gaia2018}.
In the top panel we present the additional astrometric measurement of the confirmed comoving companion TYC~8998-760-1~b which was detected at a separation of $1\farcs708\pm0\farcs003$ and a position angle of $212\fdg1\pm0\fdg1$ on the night of 2020 February 16.
The bottom panel displays the relative astrometric offsets that we measured for background contaminants within the SPHERE/IRDIS field of view.
Whereas TYC~8998-760-1~b shows no relative motion with respect to the primary within the measurement uncertainties, the background data points clearly follow the expected trajectory of a static object at infinity as indicated by the blue dashed line.
Minor deviations from this trajectory indicate intrinsic nonzero proper motions of these background objects, the measured motions, however, clearly disfavor any bound orbits for these contaminants.
As presented in the middle panel of Fig.~\ref{fig:proper_motion_diagram}, the relative proper motion of TYC~8998-760-1~c is highly inconsistent with the expected movement of a static background object.
Analogously to TYC~8998-760-1~b (top panel) its relative motion with respect to the primary is close to zero within the provided uncertainties and the measurements from 2017 July 5 and 2019 March 23 are significantly distinct from the cloud of background objects for the corresponding reference epochs.
This is in good agreement with the infinitesimal amount of orbital motion expected for an object at \rev{a projected physical separation} of 320\,au.

\subsection{Photometric analysis}
\label{subsec:photometric_analysis}

\begin{figure}
\plotone{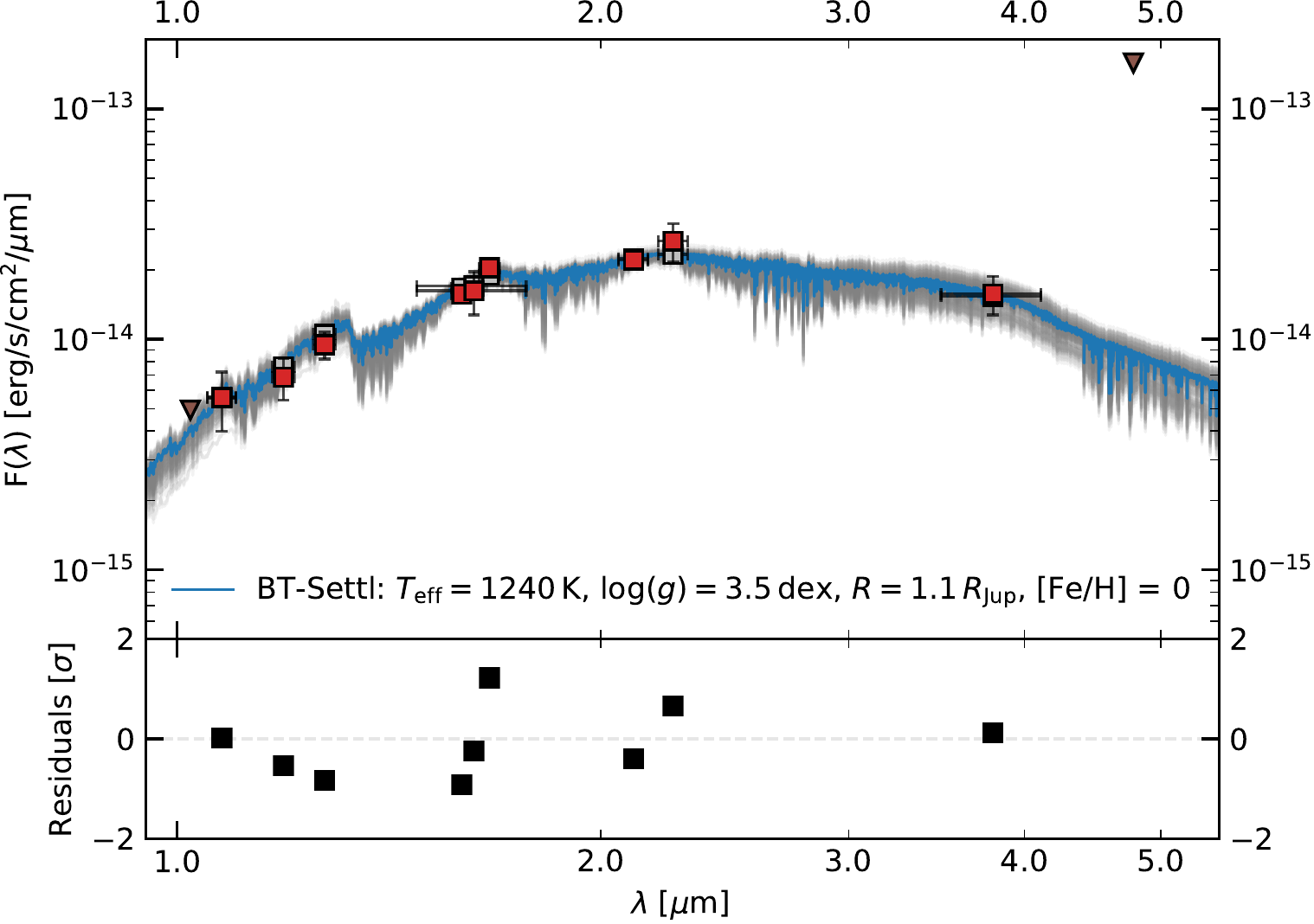}
\caption{
SED of TYC~8998-760-1.
The red squares indicate the photometric measurements we extracted from SPHERE and NACO data \rev{and the brown triangles are $5\sigma$ upper limits for bandpasses with a nondetection of the companion.
The blue line represents the median of the posterior distributions from our MCMC fitting routine} and the gray squares indicate the evaluation of this model in the SPHERE and NACO bandpasses.
We show 100 randomly drawn models from our MCMC posterior distribution (gray curves) and in the bottom panel the residuals of the \rev{posterior-median} model and the measured photometry are plotted.
}
\label{fig:sed_companion_2}
\end{figure}

To corroborate the companion status and to further characterize TYC~8998-760-1~c, we analyzed its spectral energy distribution (SED) that we constructed from the SPHERE and NACO detections ranging from $Y3$ to $L'$ band.
The $Y2$ and $M'$ data imposed additional upper limits to the SED.
As described in \citet{bohn2020a} we extracted the companion flux in the SPHERE filters by aperture photometry, choosing an aperture size equivalent to the PSF FWHM of the corresponding filter.
The magnitude contrast with respect to the primary is evaluated using the noncoronagraphic flux images that were acquired alongside the observations.
As we performed a PCA-based PSF subtraction for the reduction of the NACO $L'$ data, we extracted the magnitude of the companion by injection of negative artificial companions that were generated from the unsaturated stellar PSF in each individual frame.
This analysis was performed with the \texttt{SimplexMinimizationModule} of PynPoint that is iteratively minimizing the absolute value norm within a circular aperture around the estimated position of the companion \citep{wertz2017} using a simplex-based Nelder--Mead optimization algorithm \citep{nelder1965}.
The upper limits for $Y3$ and $M'$ bands were calculated as the \rev{5$\sigma$} detection limits at the position of the companion.
The extracted flux values are presented in Table~\ref{tbl:companion_photometry} and visualized in Fig.~\ref{fig:sed_companion_2}.
\begin{deluxetable}{@{}llll@{}}
\tablecaption{
Photometry of TYC~8998-760-1~c and Its Host.
}
\label{tbl:companion_photometry}
\tablewidth{0pt}
\tablehead{
Filter & Magnitude star & $\Delta$Mag & Flux companion\\
 & (mag) & (mag) & ($\mathrm{erg}\,\mathrm{s}^{-1}\,\mathrm{cm}^{-2}\,\mathrm{\mu m}^{-1}$)
 }
\startdata
\textit{Y2} & $9.47$ & $>13.22$ & $<0.49\times10^{-14}$\\ 
\textit{Y3} & 9.36 & $13.01\pm0.31$ & $(0.56\pm0.16)\times10^{-14}$\\
\textit{J2} & 9.13 & $12.68\pm0.22$ & $(0.69\pm0.14)\times10^{-14}$\\
\textit{J3} & 8.92 & $12.25\pm0.15$ & $(0.95\pm0.13)\times10^{-14}$\\
\textit{H2} & 8.46 & $11.32\pm0.08$ & $(1.57\pm0.11)\times10^{-14}$\\
\textit{H} & 8.44 & $11.25\pm0.23$ & $(1.62\pm0.34)\times10^{-14}$\\
\textit{H3} & 8.36 & $10.96\pm0.06$ & $(2.04\pm0.12)\times10^{-14}$\\
\textit{K1} & 8.31 & $10.03\pm0.04$ & $(2.21\pm0.09)\times10^{-14}$\\
\textit{K2} & 8.28 & $9.57\pm0.09$ & $(2.67\pm0.51)\times10^{-14}$\\
\textit{L'} & 8.27 & $8.02\pm0.21$ & $(1.58\pm0.30)\times10^{-14}$\\
\textit{M'} & $8.36$  & $>4.45$ & $<15.83\times10^{-14}$\\ 
\enddata
\tablecomments{
\rev{
We present \rev{$5\sigma$} upper limits of the companion flux in the $Y2$ and $M'$ bands.
The broadband $H$ data is reported for the night of 2020 February 16, which is superior to the data collected on 2017 July 5 due to the longer integration time and better weather conditions.
}
}
\end{deluxetable}

To assess the planetary parameters of TYC~8998-760-1~c we fitted the photometric data points with a grid of BT-Settl models \citep{allard2012} that we evaluated in the corresponding bandpasses.
We restricted this analysis to models with effective temperatures from 500\,K to 2000\,K and surface gravities ranging from 3.5\,dex to 5.5\,dex with grid spacings of 100\,K and 0.5\,dex, respectively.
\rev{
In accordance with Sco--Cen membership, only models with solar metallicity were considered for this analysis.
}
Furthermore, we assumed a negligible extinction in agreement with SED modeling of the primary as described in \citet{bohn2020a}.
\revtwo{
To facilitate model evaluation at intermediate temperatures and surface gravities we linearly interpolated the original data grid.
}

\rev{
The planetary properties were inferred by a Bayesian parameter study using the affine-invariant Markov chain Monte Carlo (MCMC) ensemble sampler implemented in the \texttt{emcee} python module \citep{foreman-mackey2013}.
The fitted parameters were the companion's effective temperature $T_\mathrm{eff}$, surface gravity $\log\left(g\right)$, and radius $R$.
Due to the negligible uncertainties in system parallax, we set the distance to a fixed value of 94.6\,pc. 
The planet luminosity for any realization of $T_\mathrm{eff}$, $\log\left(g\right)$, and $R$ was inferred from the integrated flux of the corresponding BT-Settl model, considering the previously fixed system distance.
Our MCMC implementation used uniform priors for each of the input parameters, sampling $T_\mathrm{eff}$ and $\log\left(g\right)$ over the full range of interpolated BT-Settl models and allowing for planet radii between $0.5\,R_\mathrm{Jup}$ and $5\,R_\mathrm{Jup}$.
We used a Gaussian likelihood function for the measured photometry of the companion and additionally required that the likelihood decreases to zero in case the flux in $Y$ or $M'$ bands exceeds the corresponding $5\,\sigma$ limits.
We set up an MCMC sampler with 100 walkers and 10,000 steps each for the SED fit of the companion.
Based on the derived autocorrelation times of approximately 100 iterations, we discarded the first 500 steps of the chains as burn-in phase and continued using only every twentieth step of the remaining data, which resulted in 47,500 individual posterior samples.
}

\rev{
The SED of TYC~8998-760-1~c and resulting models from our MCMC fitting procedure are presented in Fig.~\ref{fig:sed_companion_2}.
}
\rev{
From this analysis we derived estimates of $T_\mathrm{eff}=1240^{+160}_{-170}$\,K, $\log\left(g\right)=3.51^{+0.02}_{-0.01}$\,dex, $R_\mathrm{p}=1.1^{+0.6}_{-0.3}\,R_\mathrm{Jup}$, and $\log\left(L/L_\sun\right)=-4.65^{+0.05}_{-0.08}$ as the 95\,\% confidence intervals around the median of the posterior distributions\footnote{
\rev{
The full posterior distributions of this analysis and the correlations between the fitted parameters are presented in Appendix~\ref{sec:posterior_sed_fit}.
}
}.
The uncertainties derived for the surface gravity appear underestimated, as photometric measurements alone cannot precisely constrain this parameter.
We thus adopted the spacing of the original model grid of 0.5\,dex as the reported uncertainty in the planet's surface gravity henceforth.
Future measurements at higher spectral resolution are required though to place tighter constraints to this parameter.
}

To convert the derived properties to a planetary mass, we evaluated effective temperature and luminosity individually with BT-Settl isochrones at the system age of $16.7\pm1.4$\,Myr.
This yielded masses of $7.0^{+2.1}_{-1.9}\,M_\mathrm{Jup}$ and $5.5^{+0.6}_{-0.7}\,M_\mathrm{Jup}$ for both parameters, respectively.
\rev{
The planet luminosity is usually less model dependent than the derived effective temperature \citep[e.g.,][]{bonnefoy2016}, which is apparent in the uncertainties of both mass estimates.
}
We thus adopted a final mass estimate of $6\pm1\,M_\mathrm{Jup}$ for TYC~8998-760-1~c \rev{as the weighted average of both measurements}.
This is equivalent to a mass ratio of $q=0.57\pm0.10\,\%$ with respect to the primary.
Fitting the $Y$ to $K$ band data with several empirical spectra of substellar objects from \citet{chiu2006} showed best compatibility with a spectral type of L7.5.

\begin{figure}
\plotone{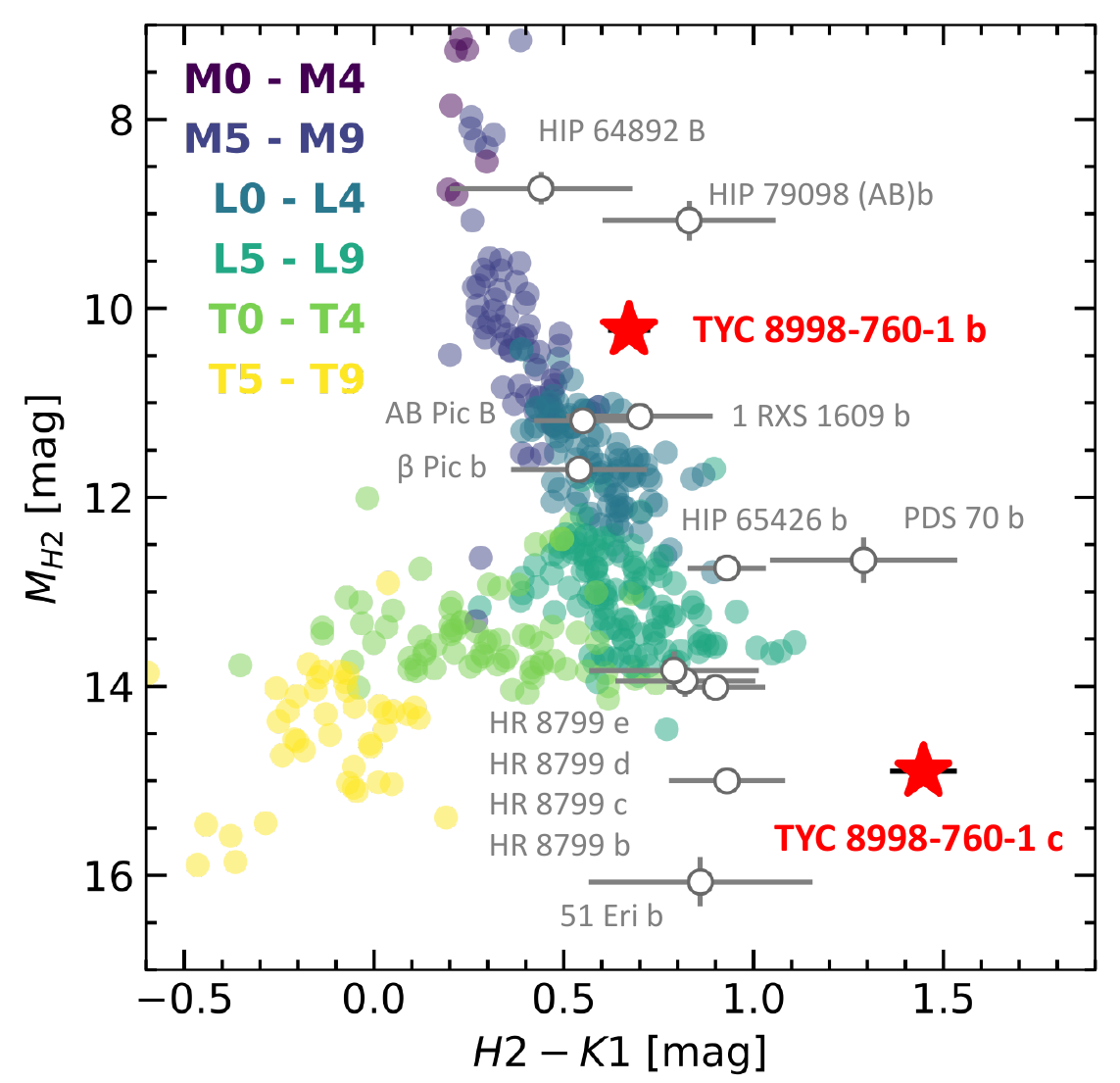}
\caption{
Color--magnitude diagram for TYC~8998-760-1~b and c.
The two objects of interest are highlighted by the red stars.
The colored, filled circles indicate the evolutionary sequence of field dwarfs of spectral class M to T and the white circles represent confirmed directly imaged companions.
}
\label{fig:color_magnitude_diagram}
\end{figure}

We further evaluated the colors of both companions with respect to field brown dwarfs and known directly imaged companions.
This analysis is presented within the color-magnitude diagram in Fig~\ref{fig:color_magnitude_diagram}.
To compile the sample of field M, L, and T dwarfs we used data provided by the NIRSPEC Brown Dwarf Spectroscopic Survey \citep{mclean2003,mclean2007}, the IRTF Spectral library \citep{rayner2009,cushing2005}, the L and T dwarf data archive \citet{knapp2004,golimowski2004,chiu2006}, and the SpeX Prism Libraries \citep{burgasser2010,gelino2010,burgasser2007,siegler2007,reid2006,kirkpatrick2006,cruz2004,burgasser2006,mcelwain2006,sheppard2009,looper2007,burgasser2008,looper2010,muench2007,dhital2011,kirkpatrick2010,burgasser2004}, using distances from Gaia DR2 \citep{gaia2018,bailerjones18}, the Brown Dwarf Kinematics Project \citep{faherty2009}, and the Pan-STARRS1 3$\pi$ Survey \citep{best2018}. 
The photometry of the directly imaged companions were adopted from \citet{chauvin2005,lafreniere2008,bonnefoy2011,currie2013,zurlo2016,samland2017,chauvin2017b,keppler2018,muller2018,cheetham2019,janson2019}.
TYC~8998-760-1~b and c are both considerably redder than the evolutionary sequence of field brown dwarfs, which is another strong indicator of their youth and low surface gravity.
TYC~8998-760-1~c is located close to the L/T transition \rev{but substantially redder than field dwarf equivalents of similar spectral type.
Indeed, it is the reddest object among the directly imaged, substellar companions that are presented in Fig.~\ref{fig:color_magnitude_diagram}.
}

\subsection{Dynamical stability}
\label{subsec:dynamical_stability}

We model the system using {\tt Rebound} and the {\tt WHFast} integrator \citep{rebound, whfast}.
We assume semi-major axes of planets b and c to be $160$ and $320$\,au respectively, and we place both planets at apastron.
For various values of the eccentricity of the planets we then calculate the chaos indicator as the mean exponential growth factor of nearby orbits \citep[MEGNO;][]{cincotta2003,megno} for the system, integrating it for its current lifetime and up to $1$\,Gyr to check its long-term stability.
We find that for orbits with low eccentricity ($e \lessapprox 0.1$) for both planets, the system is stable on gigayear timescales.
For larger eccentricities, the system is chaotic and likely to experience dynamical interaction between the planets, implying that either the planets formed in-situ or that they were ejected from the system by an unseen third companion.

\section{Conclusions}
\label{sec:conclusions}

We report the detection of TYC~8998-760-1~c: 
a second, planetary-mass companion to the solar-type Sco--Cen member TYC~8998-760-1, making this the first directly imaged system around a star of approximately $1\,M_\sun$.
From the astrometry of the object, we derived a \rev{projected physical separation} of $320$\,au.
SED analysis of broadband photometric data sampled from $Y$ to $L'$ band constrains an effective temperature of $T_\mathrm{eff}=1240^{+160}_{-170}$\,K, a surface gravity $\log\left(g\right)=3.5\pm0.5$\,dex, a planet radius of $R_\mathrm{p}=1.1^{+0.6}_{-0.3}\,R_\mathrm{Jup}$, a luminosity of $\log\left(L/L_\sun\right)=-4.65^{+0.05}_{-0.08}$, and a spectral type of L7.5.
Evaluation of BT-Settl isochrones at the system age of $16.7\pm1.4$\,Myr yielded a planet mass of $6\pm1\,M_\mathrm{Jup}$, which is consistent with a mass ratio of $q=0.57\pm0.10\,\%$ with regard to the primary.
This is in very good agreement with the color-magnitude analysis of the system that ranks TYC~8998-760-1~c as an object that is close to the L/T transition, yet much redder than field objects of the same spectral type.
\rev{
Comparison to other well-characterized, substellar companions shows that TYC~8998-760-1~c is indeed the reddest among these objects.
}
Using dynamical modeling of the system, we find that the system is stable on gigayear timescales only for near-circular orbits, with eccentric orbits becoming chaotic on timescales comparable to the system's lifetime.

TYC~8998-760-1 is a prime system to further study the dynamical and chemical properties of two coeval, gravitationally bound, gas giant planets.
Continuous astrometric monitoring will constrain the orbital solutions for both companions and thus enable testing of potential formation scenarios.
Due to the wide separations of both companions, contaminating flux from the primary is negligible, so spectral characterization at high resolution is easily accessible to determine rotational periods and molecular abundances in the planetary atmospheres \citep[e.g.][]{snellen2014}.
Multiwavelength photometric variability monitoring with space-based observatories such as the Hubble
space telescope \citep[e.g.][]{zhou2016,biller2018} and the James Webb Space Telescope (JWST) will facilitate studies of the vertical cloud structures in these Jovian companions.
Even mid-infrared spectroscopy with JWST/MIRI will be feasible to provide benchmark spectra for theoretical atmosphere models of young, substellar companions at wavelengths longer than 5 microns.

\acknowledgments

\rev{
We thank the anonymous referee for the valuable feedback that helped improve the quality of the manuscript.
}
The research of A.J.B. and F.S. leading to these results has received funding from the European Research Council under ERC Starting Grant agreement 678194 (FALCONER).
S.R. acknowledges funding from the STFC Consolidated Grant ST/R000395/1.
Part of this research was carried out at the Jet Propulsion Laboratory, California Institute of Technology, under a contract with the National Aeronautics and Space Administration.
M.R. acknowledges support from the Fonds Wetenschappelijk Onderzoek (FWO, Research Foundation Flanders) under project ID G0B3818N.
This research has made use of the SIMBAD database, operated at CDS, Strasbourg, France \citep{Wenger2000}.
This publication makes use of VOSA, developed under the Spanish Virtual Observatory project supported by the Spanish MINECO through grant AyA2017-84089.
VOSA has been partially updated by using funding from the European Union's Horizon 2020 Research and Innovation Programme, under Grant Agreement No. 776403 (EXOPLANETS-A).

To achieve the scientific results presented in this article we made use of the \emph{Python} programming language\footnote{Python Software Foundation, \url{https://www.python.org/}}, especially the \emph{SciPy} \citep{virtanen2020}, \emph{NumPy} \citep{numpy}, \emph{Matplotlib} \citep{Matplotlib}, \emph{emcee} \citep{foreman-mackey2013}, \emph{scikit-image} \citep{scikit-image}, \emph{scikit-learn} \citep{scikit-learn}, \emph{photutils} \citep{photutils}, and \emph{astropy} \citep{astropy_1,astropy_2} packages.

We performed simulations using \emph{Rebound} \citep{rebound} and \emph{AMUSE} \citep{amuse}.

%

\vspace{5mm}
\facilities{
ESO/VLT/SPHERE
ESO/VLT/NACO
}


\software{
\emph{SciPy} \citep{virtanen2020}, 
\emph{NumPy} \citep{numpy}, 
\emph{Matplotlib} \citep{Matplotlib}, 
\emph{emcee} \citep{foreman-mackey2013},
\emph{scikit-image} \citep{scikit-image}, 
\emph{scikit-learn} \citep{scikit-learn}, 
\emph{photutils} \citep{photutils}, 
\emph{astropy} \citep{astropy_1,astropy_2},
\emph{Rebound} \citep{rebound}
and \emph{AMUSE} \citep{amuse}
          }



\appendix

\section{Observational setup and conditions}
\label{sec:observations_setup_and_conditions}

The setup that was used for each observation and the weather conditions during data collection are presented in Table~\ref{tbl:observations}.

\begin{deluxetable*}{@{}llllllllll@{}}
\label{tbl:observations}
\tablecaption{High-contrast observations of TYC~8998-760-1.}
\tablewidth{0pt}
\tablehead{
Observation date & Instrument & Mode & Filter & FWHM & NEXP$\times$NDIT$\times$DIT & $\Delta\pi$ & $\langle\omega\rangle$ & $\langle X\rangle$ & $\langle\tau_0\rangle$ \\
(yyyy-mm-dd) & & & & (mas) & (1$\times$1$\times$s) & (\degr) & (\arcsec) & & (ms)
}
\startdata
2017-07-05 & SPHERE & CI & $H$ & 52.3 & 4$\times$1$\times$32 & 0.50 & 1.22 & 1.52 & 2.90 \\
2019-03-23 & SPHERE & DBI & $Y23$ & 37.2 / 37.9 & 4$\times$3$\times$64 & 3.84 & 0.41 & 1.38 & 9.30 \\
2019-03-23 & SPHERE & DBI & $J23$ & 40.1 / 41.8 & 4$\times$3$\times$64 & 3.72 & 0.40 & 1.41 & 10.75\\
2019-03-23 & SPHERE & DBI & $H23$ & 47.5 / 49.5 & 4$\times$3$\times$64 & 3.60& 0.43 & 1.44 & 10.83\\
2019-03-23 & SPHERE & DBI & $K12$ & 60.2 / 63.6 & 4$\times$3$\times$64 & 3.45 & 0.53 & 1.49 & 8.75\\
2019-05-18 & NACO & CI & $L'$ & 125.0 & 30$\times$600$\times$0.2 & 22.99 & 0.88 & 1.32 & 2.32\\
2019-06-03 & NACO & CI & $M'$ & 131.6 & 112$\times$900$\times$0.045 & 50.15 & 0.78 & 1.33 & 3.69\\
2020-02-16 & SPHERE & DPI & $H$ & 50.5 & 16$\times$4$\times$32 & 13.05 & 0.67 & 1.32 & 9.15\\
\enddata
\tablecomments{
The applied mode is either classical imaging (CI) with a broadband filter, dual-band imaging (DBI) with two intermediate band filters simultaneously, or dual-polarization imaging (DPI).
FWHM denotes the full width at half maximum that we measure from the average of the noncoronagraphic flux images that are collected for each filter. For NACO data these are equivalent to the science exposures of the star.
NEXP describes the number of exposures, NDIT is the number of subintegrations per exposure and DIT is the detector integration time of an individual subintegration.
$\Delta\pi$ denotes the amount of parallactic rotation during the observation and $\langle\omega\rangle$, $\langle X\rangle$, and $\langle\tau_0\rangle$ represent the average seeing, airmass, and coherence time, respectively.
}
\end{deluxetable*}

\section{Data reduction}
\label{sec:app_data_reduction}

\subsection{SPHERE data}
\label{subesc:app_data_reduction_sphere}

As both companions are located outside the stellar PSF halo, we did not perform any advanced post-processing for the SPHERE data:
all frames were centered and derotated accounting for the parallactic rotation of the field.
We used the standard astrometric calibration for SPHERE/IRDIS with a true north offset of $-1\fdg75\pm0\fdg08$ and plate scales varying from $12.250\pm0.010$\,mas per pixel to $12.283\pm0.010$\,mas per pixel for the applied filters as described in \citet{maire2016}.

\subsection{NACO data}
\label{subesc:app_data_reduction_naco}

As the NACO observations were optimized for the characterization of TYC~8998-760-1~b, we had to reject large fractions of the original datasets as described in Table~\ref{tbl:observations}, because TYC~8998-760-1~c was located outside the detector window for these frames.
After additional frame selection to reject frames with bad AO correction, approximately 30\,\% and 15\,\% of the full data was remaining for $L'$ and $M'$ data, respectively.
As the amount of parallactic rotation in the data was sufficient, we performed a PSF subtraction based on principal component analysis \citep[PCA;][]{pynpoint,soummer2012}.
For both $L'$ and $M'$ data, we fitted and subtracted one principal component from the images.
This was optimizing the signal-to-noise ratio of TYC~8998-760-1~c for the $L'$ data and it provided the best upper limit for the $M'$ data at the position of the companion.

\section{Signal-to-noise Assessment}
\label{sec:signal_to_noise}

\begin{figure*}
\includegraphics[width=\textwidth]{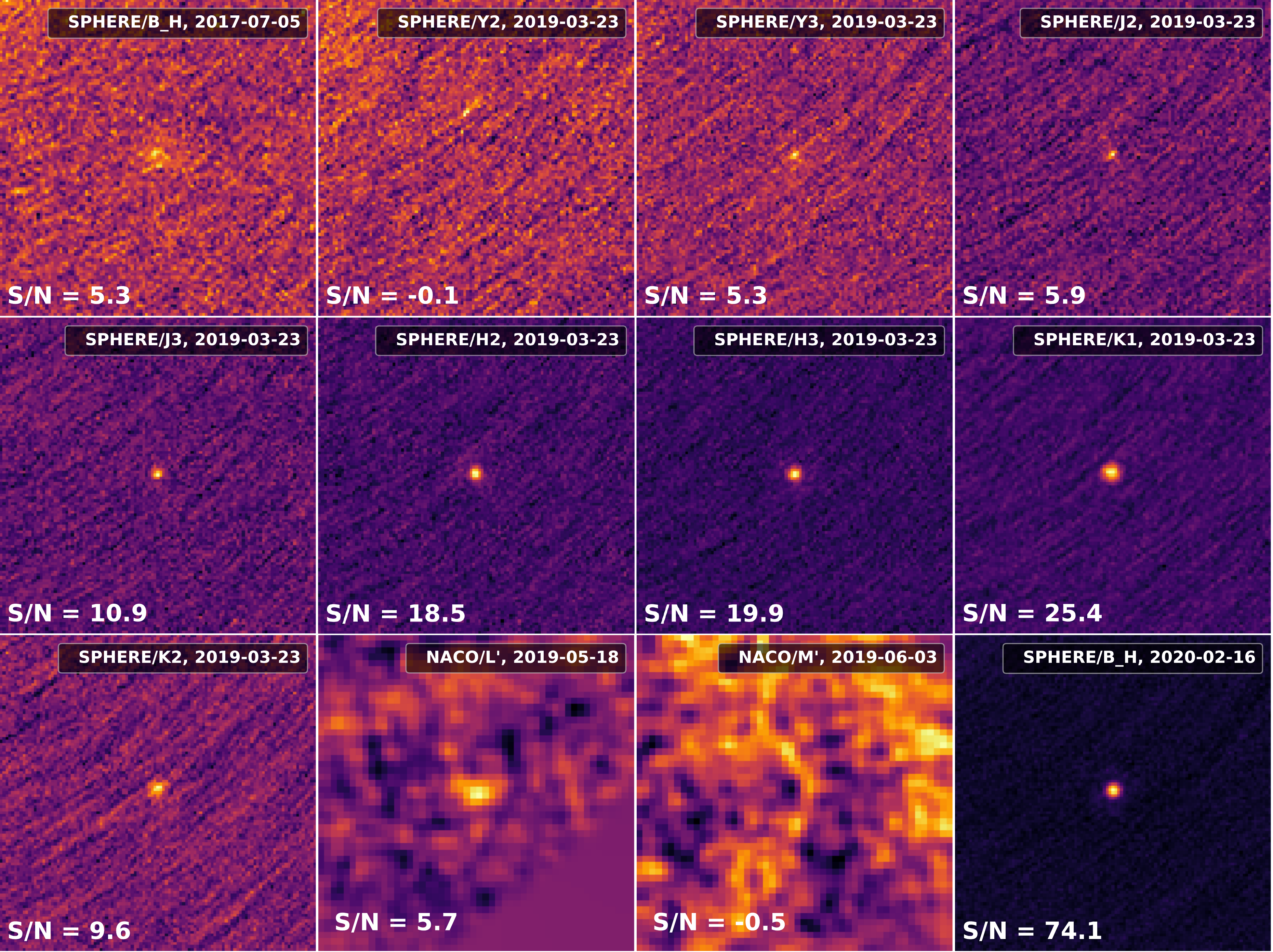}
\caption{
\rev{
Signal-to-noise ratio assessment of TYC~8998-760-1~c.
We show a cutout of the final images for all filters and epochs.
The signal-to-noise ratios of the companion were measured with aperture photometry and the resulting values are presented in the lower left of each panel.
Each image is presented on an individual linear color scale that is normalized with respect to the maximum and minimum flux value within the image cutout.
}
}
\label{fig:sn_assessment}
\end{figure*}

\rev{
To assess the significance of the detection of TYC~8998-760-1~c for each individual epoch and filter, we measured the signal-to-noise ratio of the companion in the processed images.
We evaluated the signal flux in a circular aperture placed at the previously determined position of the companion for the corresponding filter (see Sect.~\ref{subsec:astrometric_analysis}).
For bandpasses in which the companion is not detected (i.e. $Y2$ band on the night of 2019 March 23 and $M'$ band on the night of 2019 June 3), we used the astrometric position of the $K1$ data from 2019 March 23 instead.
The aperture radius was chosen as the FWHM of the unsaturated flux PSF of the corresponding filter as reported in Table~\ref{tbl:observations}.
To measure the noise, we distributed circular apertures of the same size radially around the star at the same radial separation as the companion.
We calculated the integrated flux within each of the background apertures and subtracted the average of these measurements from the integrated signal flux in the science aperture.
The noise was computed as the standard deviation of the integrated fluxes from the background apertures, following the description of \citet{mawet2014}.
The resulting signal-to-noise ratios are presented in Fig~\ref{fig:sn_assessment}.
Besides nondetections in the $Y2$ and the $M'$ data, we measure the flux of TYC~8998-760-1~c with a signal-to-noise ratio greater than 5.
}

\section{Posterior distributions of SED fit}
\label{sec:posterior_sed_fit}

\begin{figure*}
\includegraphics[width=\textwidth]{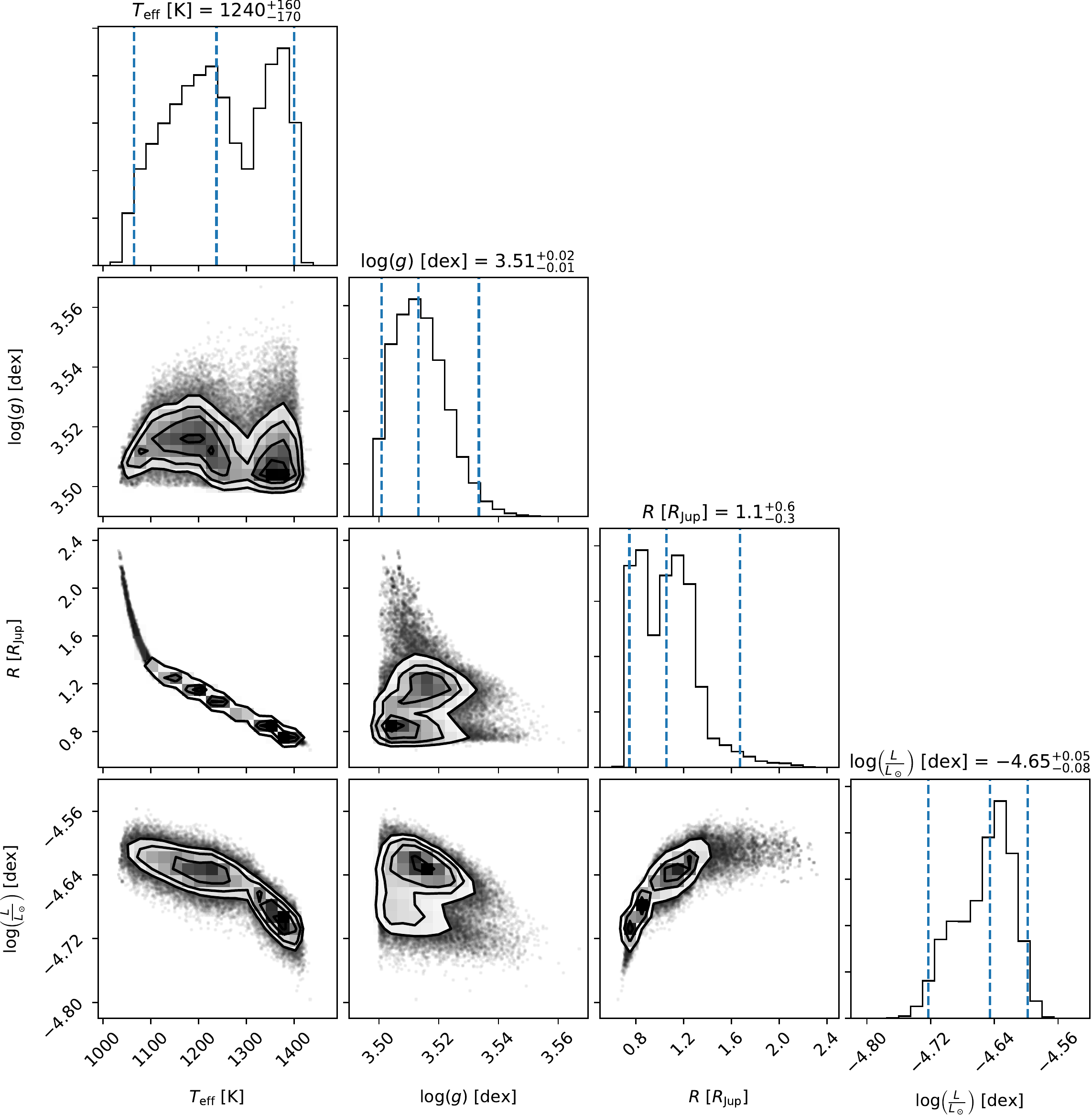}
\caption{
\rev{
Posterior distributions of the MCMC fitting procedure to the photometric SED of TYC~8998-760-1~c.
The input parameters of the fit were effective temperature $T_\mathrm{eff}$, surface gravity $\log\left(g\right)$, and object radius $R$.
We further show the resulting planet luminosities that can be derived from the three input parameters and the system distance.
The dashed blue lines in the marginalized distributions present the 2.5\,\%, 50\,\%, and 97.5\,\% quantiles and the title of the corresponding diagram indicates the 95\,\% confidence interval around the median, derived from these quantities.
}
}
\label{fig:sed_mcmc_corner}
\end{figure*}

\rev{
We present the full parameter space of posterior samples from our SED fit of TYC~8998-760-1~c in Fig.~\ref{fig:sed_mcmc_corner}.
\revtwo{
Due to the linear interpolation of the model grid prior to the MCMC fitting routine, each parameter is sampled continuously within the predefined intervals. 
}
The upper three panels of the corner plot show the correlations between the three input parameters $T_\mathrm{eff}$, $\log\left(g\right)$, and $R$.
Furthermore, we present the corresponding planet luminosities that are derived from these input parameters and the system distance in the bottom panel of the figure.
The posterior distributions show two families of solutions with effective temperatures of approximately 1225\,K and 1375\,K and associated planet radii of $1.2\,R_\mathrm{Jup}$ and $0.8\,R_\mathrm{Jup}$, respectively.
\revtwo{
Even though the latter family of solutions is slightly disfavored due to the corresponding planet radius of $0.8\,R_\mathrm{Jup}$ -- which is smaller than theoretical predictions and empirical constraints for an object of this age and mass \citep[e.g.,][]{chabrier2009,mordasini2012} --
}
we report the 95\,\% confidence intervals around the medians of the distributions as a conservative estimate of the planetary properties.
This estimate can certainly be refined by future studies at higher spectral resolution.
}

\section{Dynamical modeling}
\label{sec:dynamical_modeling}

\begin{figure}
\includegraphics[width=\columnwidth]{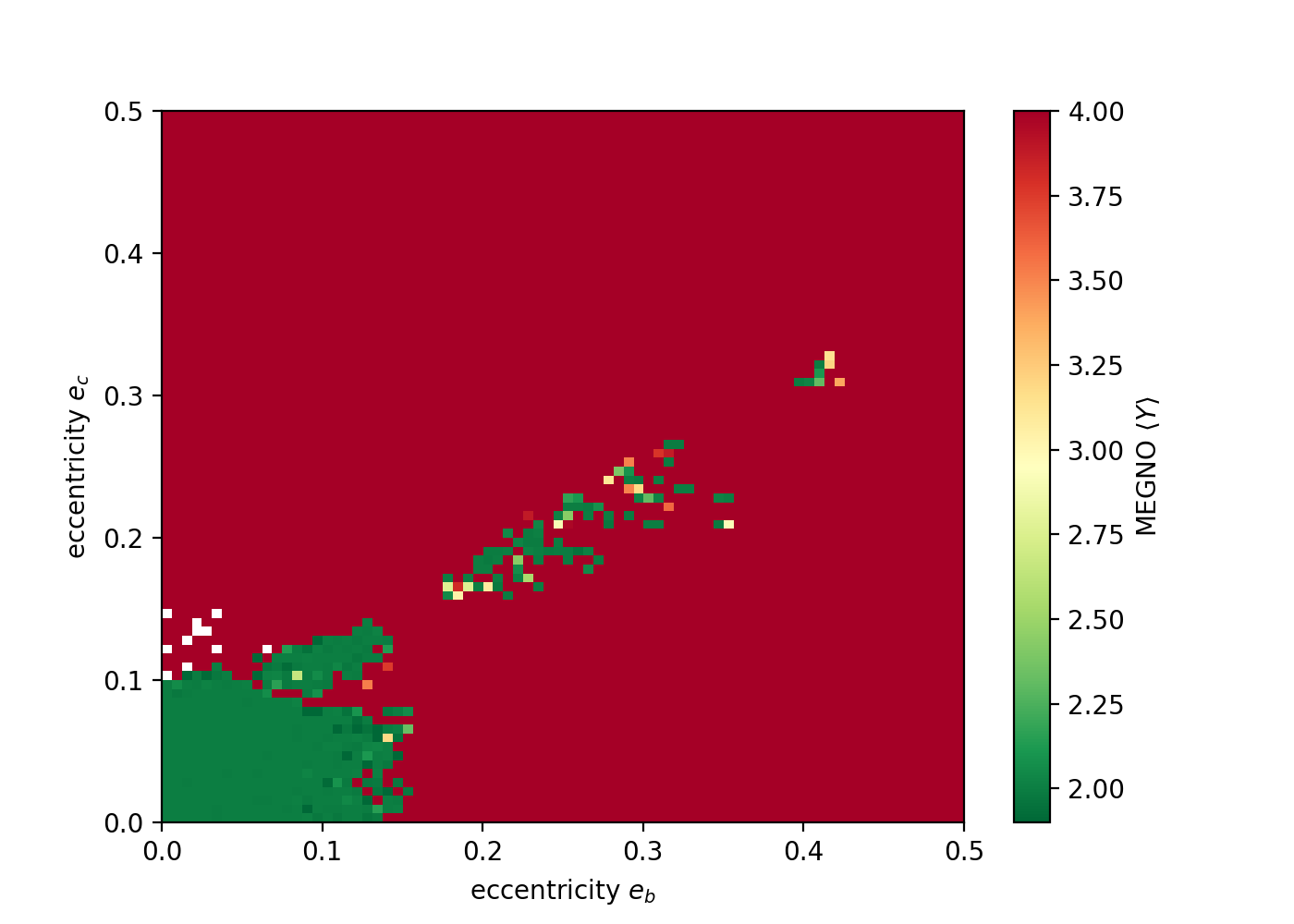}
\caption{
MEGNO value for the system after $17$\,Myr, for different eccentricities of planets $b$ and $c$.
A value $> 2$ indicates a chaotic system.
}
\label{fig:dyn_17Myr}
\end{figure}

\begin{figure}
\includegraphics[width=\columnwidth]{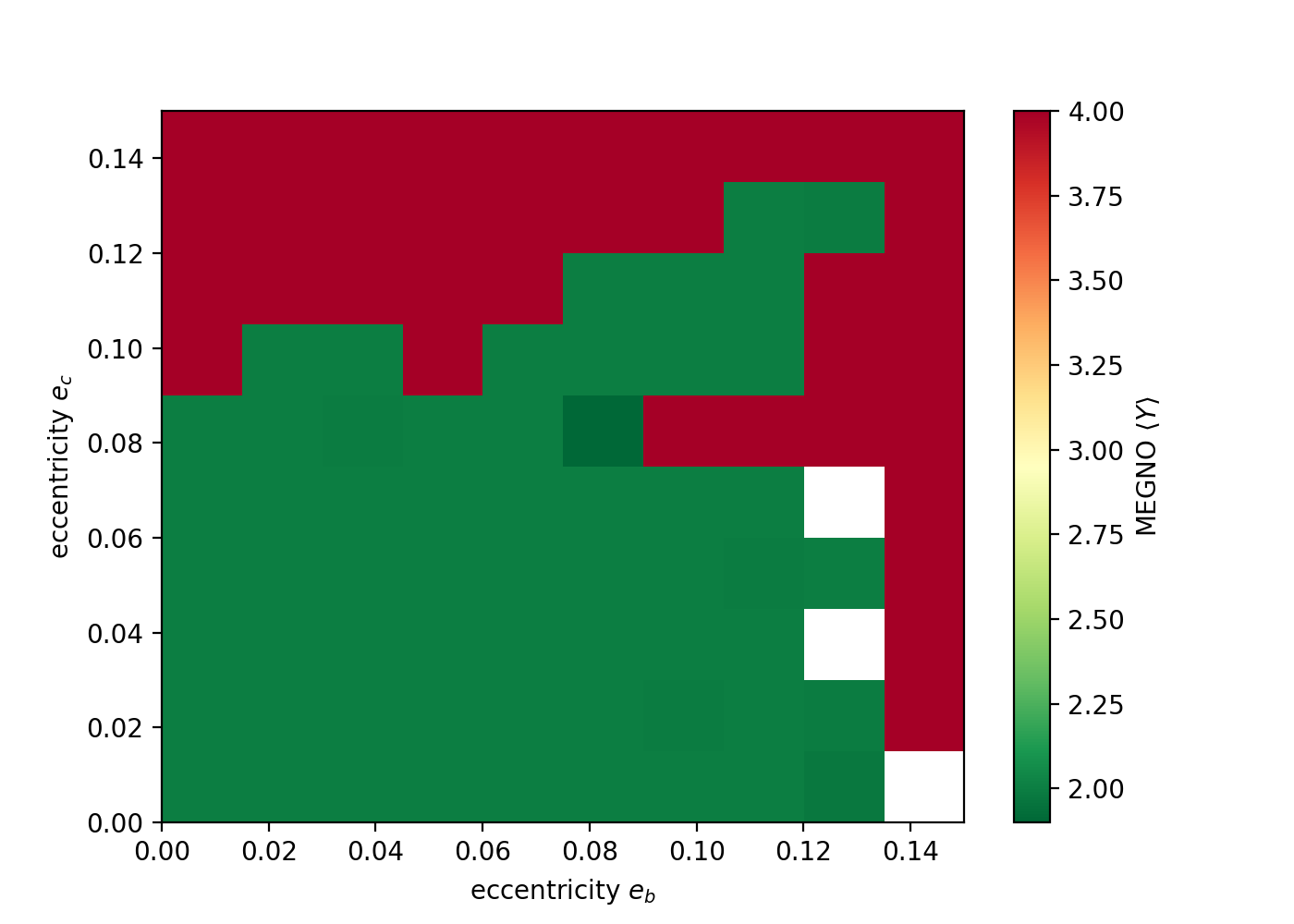}
\caption{
As Fig.~\ref{fig:dyn_17Myr}, but on a $1$\,Gyr timescale. 
We have not plotted orbits with $e > 0.15$, as they are all chaotic on this timescale.
}
\label{fig:dyn_1Gyr}
\end{figure}

In Figures~\ref{fig:dyn_17Myr} and \ref{fig:dyn_1Gyr}, we show the MEGNO values for systems after simulating them for $17$\,Myr and $1$\,Gyr, respectively (see subsection~\ref{subsec:dynamical_stability}).
A MEGNO value $> 2$ indicates a chaotic system, for which we cannot accurately predict the orbits on these timescales.

\bibliography{mybib}{}

\begin{thebibliography}{}
\expandafter\ifx\csname natexlab\endcsname\relax\def\natexlab#1{#1}\fi
\providecommand{\url}[1]{\href{#1}{#1}}
\providecommand{\dodoi}[1]{doi:~\href{http://doi.org/#1}{\nolinkurl{#1}}}
\providecommand{\doeprint}[1]{\href{http://ascl.net/#1}{\nolinkurl{http://ascl.net/#1}}}
\providecommand{\doarXiv}[1]{\href{https://arxiv.org/abs/#1}{\nolinkurl{https://arxiv.org/abs/#1}}}

\bibitem[{{Allard} {et~al.}(2012){Allard}, {Homeier}, \&
  {Freytag}}]{allard2012}
{Allard}, F., {Homeier}, D., \& {Freytag}, B. 2012, Philosophical Transactions
  of the Royal Society of London Series A, 370, 2765,
  \dodoi{10.1098/rsta.2011.0269}

\bibitem[{{Amara} \& {Quanz}(2012)}]{pynpoint}
{Amara}, A., \& {Quanz}, S.~P. 2012, \mnras, 427, 948,
  \dodoi{10.1111/j.1365-2966.2012.21918.x}

\bibitem[{{Astropy Collaboration} {et~al.}(2013){Astropy Collaboration},
  {Robitaille}, {Tollerud}, {Greenfield}, {Droettboom}, {Bray}, {Aldcroft},
  {Davis}, {Ginsburg}, {Price-Whelan}, {Kerzendorf}, {Conley}, {Crighton},
  {Barbary}, {Muna}, {Ferguson}, {Grollier}, {Parikh}, {Nair}, {Unther},
  {Deil}, {Woillez}, {Conseil}, {Kramer}, {Turner}, {Singer}, {Fox}, {Weaver},
  {Zabalza}, {Edwards}, {Azalee Bostroem}, {Burke}, {Casey}, {Crawford},
  {Dencheva}, {Ely}, {Jenness}, {Labrie}, {Lim}, {Pierfederici}, {Pontzen},
  {Ptak}, {Refsdal}, {Servillat}, \& {Streicher}}]{astropy_1}
{Astropy Collaboration}, {Robitaille}, T.~P., {Tollerud}, E.~J., {et~al.} 2013,
  \aap, 558, A33, \dodoi{10.1051/0004-6361/201322068}

\bibitem[{{Astropy Collaboration} {et~al.}(2018){Astropy Collaboration},
  {Price-Whelan}, {Sip{\H o}cz}, {G{\"u}nther}, {Lim}, {Crawford}, {Conseil},
  {Shupe}, {Craig}, {Dencheva}, {Ginsburg}, {VanderPlas}, {Bradley},
  {P{\'e}rez-Su{\'a}rez}, {de Val-Borro}, {Aldcroft}, {Cruz}, {Robitaille},
  {Tollerud}, {Ardelean}, {Babej}, {Bach}, {Bachetti}, {Bakanov}, {Bamford},
  {Barentsen}, {Barmby}, {Baumbach}, {Berry}, {Biscani}, {Boquien}, {Bostroem},
  {Bouma}, {Brammer}, {Bray}, {Breytenbach}, {Buddelmeijer}, {Burke},
  {Calderone}, {Cano Rodr{\'{\i}}guez}, {Cara}, {Cardoso}, {Cheedella},
  {Copin}, {Corrales}, {Crichton}, {D'Avella}, {Deil}, {Depagne}, {Dietrich},
  {Donath}, {Droettboom}, {Earl}, {Erben}, {Fabbro}, {Ferreira}, {Finethy},
  {Fox}, {Garrison}, {Gibbons}, {Goldstein}, {Gommers}, {Greco}, {Greenfield},
  {Groener}, {Grollier}, {Hagen}, {Hirst}, {Homeier}, {Horton}, {Hosseinzadeh},
  {Hu}, {Hunkeler}, {Ivezi{\'c}}, {Jain}, {Jenness}, {Kanarek}, {Kendrew},
  {Kern}, {Kerzendorf}, {Khvalko}, {King}, {Kirkby}, {Kulkarni}, {Kumar},
  {Lee}, {Lenz}, {Littlefair}, {Ma}, {Macleod}, {Mastropietro}, {McCully},
  {Montagnac}, {Morris}, {Mueller}, {Mumford}, {Muna}, {Murphy}, {Nelson},
  {Nguyen}, {Ninan}, {N{\"o}the}, {Ogaz}, {Oh}, {Parejko}, {Parley}, {Pascual},
  {Patil}, {Patil}, {Plunkett}, {Prochaska}, {Rastogi}, {Reddy Janga},
  {Sabater}, {Sakurikar}, {Seifert}, {Sherbert}, {Sherwood-Taylor}, {Shih},
  {Sick}, {Silbiger}, {Singanamalla}, {Singer}, {Sladen}, {Sooley},
  {Sornarajah}, {Streicher}, {Teuben}, {Thomas}, {Tremblay}, {Turner},
  {Terr{\'o}n}, {van Kerkwijk}, {de la Vega}, {Watkins}, {Weaver}, {Whitmore},
  {Woillez}, {Zabalza}, \& {Astropy Contributors}}]{astropy_2}
{Astropy Collaboration}, {Price-Whelan}, A.~M., {Sip{\H o}cz}, B.~M., {et~al.}
  2018, \aj, 156, 123, \dodoi{10.3847/1538-3881/aabc4f}

\bibitem[{{Bailer-Jones} {et~al.}(2018){Bailer-Jones}, {Rybizki}, {Fouesneau},
  {Mantelet}, \& {Andrae}}]{bailerjones18}
{Bailer-Jones}, C.~A.~L., {Rybizki}, J., {Fouesneau}, M., {Mantelet}, G., \&
  {Andrae}, R. 2018, \aj, 156, 58, \dodoi{10.3847/1538-3881/aacb21}

\bibitem[{{Best} {et~al.}(2018){Best}, {Magnier}, {Liu}, {Aller}, {Zhang},
  {Burgett}, {Chambers}, {Draper}, {Flewelling}, {Kaiser}, {Kudritzki},
  {Metcalfe}, {Tonry}, {Wainscoat}, \& {Waters}}]{best2018}
{Best}, W. M.~J., {Magnier}, E.~A., {Liu}, M.~C., {et~al.} 2018, \apjs, 234, 1,
  \dodoi{10.3847/1538-4365/aa9982}

\bibitem[{{Beuzit} {et~al.}(2019){Beuzit}, {Vigan}, {Mouillet}, {Dohlen},
  {Gratton}, {Boccaletti}, {Sauvage}, {Schmid}, {Langlois}, {Petit},
  {Baruffolo}, {Feldt}, {Milli}, {Wahhaj}, {Abe}, {Anselmi}, {Antichi},
  {Barette}, {Baudrand}, {Baudoz}, {Bazzon}, {Bernardi}, {Blanchard}, {Brast},
  {Bruno}, {Buey}, {Carbillet}, {Carle}, {Cascone}, {Chapron}, {Charton},
  {Chauvin}, {Claudi}, {Costille}, {De Caprio}, {de Boer}, {Delboulb{\'e}},
  {Desidera}, {Dominik}, {Downing}, {Dupuis}, {Fabron}, {Fantinel}, {Farisato},
  {Feautrier}, {Fedrigo}, {Fusco}, {Gigan}, {Ginski}, {Girard}, {Giro},
  {Gisler}, {Gluck}, {Gry}, {Henning}, {Hubin}, {Hugot}, {Incorvaia}, {Jaquet},
  {Kasper}, {Lagadec}, {Lagrange}, {Le Coroller}, {Le Mignant}, {Le Ruyet},
  {Lessio}, {Lizon}, {Llored}, {Lundin}, {Madec}, {Magnard}, {Marteaud},
  {Martinez}, {Maurel}, {M{\'e}nard}, {Mesa}, {M{\"o}ller-Nilsson}, {Moulin},
  {Moutou}, {Orign{\'e}}, {Parisot}, {Pavlov}, {Perret}, {Pragt}, {Puget},
  {Rabou}, {Ramos}, {Reess}, {Rigal}, {Rochat}, {Roelfsema}, {Rousset}, {Roux},
  {Saisse}, {Salasnich}, {Santambrogio}, {Scuderi}, {Segransan}, {Sevin},
  {Siebenmorgen}, {Soenke}, {Stadler}, {Suarez}, {Tiph{\`e}ne}, {Turatto},
  {Udry}, {Vakili}, {Waters}, {Weber}, {Wildi}, {Zins}, \&
  {Zurlo}}]{Beuzit2019}
{Beuzit}, J.~L., {Vigan}, A., {Mouillet}, D., {et~al.} 2019, \aap, 631, A155,
  \dodoi{10.1051/0004-6361/201935251}

\bibitem[{{Biller} {et~al.}(2018){Biller}, {Vos}, {Buenzli}, {Allers},
  {Bonnefoy}, {Charnay}, {B{\'e}zard}, {Allard}, {Homeier}, {Bonavita},
  {Brandner}, {Crossfield}, {Dupuy}, {Henning}, {Kopytova}, {Liu},
  {Manjavacas}, \& {Schlieder}}]{biller2018}
{Biller}, B.~A., {Vos}, J., {Buenzli}, E., {et~al.} 2018, \aj, 155, 95,
  \dodoi{10.3847/1538-3881/aaa5a6}

\bibitem[{{Bohn} {et~al.}(2019){Bohn}, {Kenworthy}, {Ginski}, {Benisty}, {de
  Boer}, {Keller}, {Mamajek}, {Meshkat}, {Muro-Arena}, {Pecaut}, {Snik},
  {Wolff}, \& {Reggiani}}]{bohn2019}
{Bohn}, A.~J., {Kenworthy}, M.~A., {Ginski}, C., {et~al.} 2019, \aap, 624, A87,
  \dodoi{10.1051/0004-6361/201834523}

\bibitem[{{Bohn} {et~al.}(2020){Bohn}, {Kenworthy}, {Ginski}, {Manara},
  {Pecaut}, {de Boer}, {Keller}, {Mamajek}, {Meshkat}, {Reggiani}, {Todorov},
  \& {Snik}}]{bohn2020a}
---. 2020, \mnras, 492, 431, \dodoi{10.1093/mnras/stz3462}

\bibitem[{{Bonnefoy} {et~al.}(2011){Bonnefoy}, {Lagrange}, {Boccaletti},
  {Chauvin}, {Apai}, {Allard}, {Ehrenreich}, {Girard}, {Mouillet}, {Rouan},
  {Gratadour}, \& {Kasper}}]{bonnefoy2011}
{Bonnefoy}, M., {Lagrange}, A.~M., {Boccaletti}, A., {et~al.} 2011, \aap, 528,
  L15, \dodoi{10.1051/0004-6361/201016224}

\bibitem[{{Bonnefoy} {et~al.}(2016){Bonnefoy}, {Zurlo}, {Baudino}, {Lucas},
  {Mesa}, {Maire}, {Vigan}, {Galicher}, {Homeier}, {Marocco}, {Gratton},
  {Chauvin}, {Allard}, {Desidera}, {Kasper}, {Moutou}, {Lagrange}, {Antichi},
  {Baruffolo}, {Baudrand }, {Beuzit}, {Boccaletti}, {Cantalloube}, {Carbillet},
  {Charton}, {Claudi}, {Costille}, {Dohlen}, {Dominik}, {Fantinel},
  {Feautrier}, {Feldt}, {Fusco}, {Gigan}, {Girard}, {Gluck}, {Gry}, {Henning},
  {Janson}, {Langlois}, {Madec}, {Magnard}, {Maurel}, {Mawet}, {Meyer},
  {Milli}, {Moeller-Nilsson}, {Mouillet}, {Pavlov}, {Perret}, {Pujet}, {Quanz},
  {Rochat}, {Rousset}, {Roux}, {Salasnich}, {Salter}, {Sauvage}, {Schmid},
  {Sevin}, {Soenke}, {Stadler}, {Turatto}, {Udry}, {Vakili}, {Wahhaj}, \&
  {Wildi}}]{bonnefoy2016}
{Bonnefoy}, M., {Zurlo}, A., {Baudino}, J.~L., {et~al.} 2016, \aap, 587, A58,
  \dodoi{10.1051/0004-6361/201526906}

\bibitem[{{Bradley} {et~al.}(2016){Bradley}, {Sipocz}, {Robitaille},
  {Tollerud}, {Deil}, {Vin{\'\i}cius}, {Barbary}, {G{\"u}nther}, {Bostroem},
  {Droettboom}, {Bray}, {Bratholm}, {Pickering}, {Craig}, {Pascual}, {Greco},
  {Donath}, {Kerzendorf}, {Littlefair}, {Barentsen}, {D'Eugenio}, \&
  {Weaver}}]{photutils}
{Bradley}, L., {Sipocz}, B., {Robitaille}, T., {et~al.} 2016, {Photutils:
  Photometry tools}.
\newblock \doeprint{1609.011}

\bibitem[{{Burgasser}(2007)}]{burgasser2007}
{Burgasser}, A.~J. 2007, \apj, 659, 655, \dodoi{10.1086/511027}

\bibitem[{{Burgasser} {et~al.}(2010){Burgasser}, {Cruz}, {Cushing}, {Gelino},
  {Looper}, {Faherty}, {Kirkpatrick}, \& {Reid}}]{burgasser2010}
{Burgasser}, A.~J., {Cruz}, K.~L., {Cushing}, M., {et~al.} 2010, \apj, 710,
  1142, \dodoi{10.1088/0004-637X/710/2/1142}

\bibitem[{{Burgasser} {et~al.}(2008){Burgasser}, {Liu}, {Ireland}, {Cruz}, \&
  {Dupuy}}]{burgasser2008}
{Burgasser}, A.~J., {Liu}, M.~C., {Ireland}, M.~J., {Cruz}, K.~L., \& {Dupuy},
  T.~J. 2008, \apj, 681, 579, \dodoi{10.1086/588379}

\bibitem[{{Burgasser} \& {McElwain}(2006)}]{burgasser2006}
{Burgasser}, A.~J., \& {McElwain}, M.~W. 2006, \aj, 131, 1007,
  \dodoi{10.1086/499042}

\bibitem[{{Burgasser} {et~al.}(2004){Burgasser}, {McElwain}, {Kirkpatrick},
  {Cruz}, {Tinney}, \& {Reid}}]{burgasser2004}
{Burgasser}, A.~J., {McElwain}, M.~W., {Kirkpatrick}, J.~D., {et~al.} 2004,
  \aj, 127, 2856, \dodoi{10.1086/383549}

\bibitem[{{Chabrier} {et~al.}(2009){Chabrier}, {Baraffe}, {Leconte},
  {Gallardo}, \& {Barman}}]{chabrier2009}
{Chabrier}, G., {Baraffe}, I., {Leconte}, J., {Gallardo}, J., \& {Barman}, T.
  2009, in American Institute of Physics Conference Series, Vol. 1094, 15th
  Cambridge Workshop on Cool Stars, Stellar Systems, and the Sun, ed.
  E.~{Stempels}, 102--111, \dodoi{10.1063/1.3099078}

\bibitem[{{Chauvin} {et~al.}(2005){Chauvin}, {Lagrange}, {Zuckerman}, {Dumas},
  {Mouillet}, {Song}, {Beuzit}, {Lowrance}, \& {Bessell}}]{chauvin2005}
{Chauvin}, G., {Lagrange}, A.~M., {Zuckerman}, B., {et~al.} 2005, Astronomy and
  Astrophysics, 438, L29, \dodoi{10.1051/0004-6361:200500111}

\bibitem[{{Chauvin} {et~al.}(2017{\natexlab{a}}){Chauvin}, {Desidera},
  {Lagrange}, {Vigan}, {Feldt}, {Gratton}, {Langlois}, {Cheetham}, {Bonnefoy},
  \& {Meyer}}]{chauvin2017a}
{Chauvin}, G., {Desidera}, S., {Lagrange}, A.~M., {et~al.} 2017{\natexlab{a}},
  in SF2A-2017: Proceedings of the Annual meeting of the French Society of
  Astronomy and Astrophysics, Di

\bibitem[{{Chauvin} {et~al.}(2017{\natexlab{b}}){Chauvin}, {Desidera},
  {Lagrange}, {Vigan}, {Gratton}, {Langlois}, {Bonnefoy}, {Beuzit}, {Feldt},
  {Mouillet}, {Meyer}, {Cheetham}, {Biller}, {Boccaletti}, {D'Orazi},
  {Galicher}, {Hagelberg}, {Maire}, {Mesa}, {Olofsson}, {Samland}, {Schmidt},
  {Sissa}, {Bonavita}, {Charnay}, {Cudel}, {Daemgen}, {Delorme},
  {Janin-Potiron}, {Janson}, {Keppler}, {Le Coroller}, {Ligi}, {Marleau},
  {Messina}, {Molli{\`e}re}, {Mordasini}, {M{\"u}ller}, {Peretti}, {Perrot},
  {Rodet}, {Rouan}, {Zurlo}, {Dominik}, {Henning}, {Menard}, {Schmid},
  {Turatto}, {Udry}, {Vakili}, {Abe}, {Antichi}, {Baruffolo}, {Baudoz},
  {Baudrand}, {Blanchard}, {Bazzon}, {Buey}, {Carbillet}, {Carle}, {Charton},
  {Cascone}, {Claudi}, {Costille}, {Deboulbe}, {De Caprio}, {Dohlen},
  {Fantinel}, {Feautrier}, {Fusco}, {Gigan}, {Giro}, {Gisler}, {Gluck},
  {Hubin}, {Hugot}, {Jaquet}, {Kasper}, {Madec}, {Magnard}, {Martinez},
  {Maurel}, {Le Mignant}, {M{\"o}ller-Nilsson}, {Llored}, {Moulin},
  {Orign{\'e}}, {Pavlov}, {Perret}, {Petit}, {Pragt}, {Puget}, {Rabou},
  {Ramos}, {Rigal}, {Rochat}, {Roelfsema}, {Rousset}, {Roux}, {Salasnich},
  {Sauvage}, {Sevin}, {Soenke}, {Stadler}, {Suarez}, {Weber}, {Wildi},
  {Antoniucci}, {Augereau}, {Baudino}, {Brandner}, {Engler}, {Girard}, {Gry},
  {Kral}, {Kopytova}, {Lagadec}, {Milli}, {Moutou}, {Schlieder},
  {Szul{\'a}gyi}, {Thalmann}, \& {Wahhaj}}]{chauvin2017b}
{Chauvin}, G., {Desidera}, S., {Lagrange}, A.~M., {et~al.} 2017{\natexlab{b}},
  Astronomy and Astrophysics, 605, L9, \dodoi{10.1051/0004-6361/201731152}

\bibitem[{{Cheetham} {et~al.}(2019){Cheetham}, {Samland}, {Brems}, {Launhardt},
  {Chauvin}, {S{\'e}gransan}, {Henning}, {Quirrenbach}, {Avenhaus}, {Cugno},
  {Girard}, {Godoy}, {Kennedy}, {Maire}, {Metchev}, {M{\"u}ller}, {Musso
  Barcucci}, {Olofsson}, {Pepe}, {Quanz}, {Queloz}, {Reffert}, {Rickman}, {van
  Boekel}, {Boccaletti}, {Bonnefoy}, {Cantalloube}, {Charnay}, {Delorme},
  {Janson}, {Keppler}, {Lagrange}, {Langlois}, {Lazzoni}, {Menard}, {Mesa},
  {Meyer}, {Schmidt}, {Sissa}, {Udry}, \& {Zurlo}}]{cheetham2019}
{Cheetham}, A.~C., {Samland}, M., {Brems}, S.~S., {et~al.} 2019, \aap, 622,
  A80, \dodoi{10.1051/0004-6361/201834112}

\bibitem[{{Chen} {et~al.}(2012){Chen}, {Pecaut}, {Mamajek}, {Su}, \&
  {Bitner}}]{chen2012}
{Chen}, C.~H., {Pecaut}, M., {Mamajek}, E.~E., {Su}, K. Y.~L., \& {Bitner}, M.
  2012, \apj, 756, 133, \dodoi{10.1088/0004-637X/756/2/133}

\bibitem[{{Chiu} {et~al.}(2006){Chiu}, {Fan}, {Leggett}, {Golimowski}, {Zheng},
  {Geballe}, {Schneider}, \& {Brinkmann}}]{chiu2006}
{Chiu}, K., {Fan}, X., {Leggett}, S.~K., {et~al.} 2006, \aj, 131, 2722,
  \dodoi{10.1086/501431}

\bibitem[{{Cincotta} {et~al.}(2003){Cincotta}, {Giordano}, \&
  {Sim{\'o}}}]{cincotta2003}
{Cincotta}, P.~M., {Giordano}, C.~M., \& {Sim{\'o}}, C. 2003, Physica D
  Nonlinear Phenomena, 182, 151, \dodoi{10.1016/S0167-2789(03)00103-9}

\bibitem[{{Cruz} {et~al.}(2004){Cruz}, {Burgasser}, {Reid}, \&
  {Liebert}}]{cruz2004}
{Cruz}, K.~L., {Burgasser}, A.~J., {Reid}, I.~N., \& {Liebert}, J. 2004, \apjl,
  604, L61, \dodoi{10.1086/383415}

\bibitem[{{Currie} {et~al.}(2013){Currie}, {Burrows}, {Madhusudhan},
  {Fukagawa}, {Girard}, {Dawson}, {Murray-Clay}, {Kenyon}, {Kuchner},
  {Matsumura}, {Jayawardhana}, {Chambers}, \& {Bromley}}]{currie2013}
{Currie}, T., {Burrows}, A., {Madhusudhan}, N., {et~al.} 2013, \apj, 776, 15,
  \dodoi{10.1088/0004-637X/776/1/15}

\bibitem[{{Cushing} {et~al.}(2005){Cushing}, {Rayner}, \&
  {Vacca}}]{cushing2005}
{Cushing}, M.~C., {Rayner}, J.~T., \& {Vacca}, W.~D. 2005, \apj, 623, 1115,
  \dodoi{10.1086/428040}

\bibitem[{{de Boer} {et~al.}(2020){de Boer}, {Langlois}, {van Holstein},
  {Girard}, {Mouillet}, {Vigan}, {Dohlen}, {Snik}, {Keller}, {Ginski}, {Stam},
  {Milli}, {Wahhaj}, {Kasper}, {Schmid}, {Rabou}, {Gluck}, {Hugot}, {Perret},
  {Martinez}, {Weber}, {Pragt}, {Sauvage}, {Boccaletti}, {Le Coroller},
  {Dominik}, {Henning}, {Lagadec}, {M{\'e}nard}, {Turatto}, {Udry}, {Chauvin},
  {Feldt}, \& {Beuzit}}]{deBoer2020}
{de Boer}, J., {Langlois}, M., {van Holstein}, R.~G., {et~al.} 2020, \aap, 633,
  A63, \dodoi{10.1051/0004-6361/201834989}

\bibitem[{{de Zeeuw} {et~al.}(1999){de Zeeuw}, {Hoogerwerf}, {de Bruijne},
  {Brown}, \& {Blaauw}}]{dezeeuw1999}
{de Zeeuw}, P.~T., {Hoogerwerf}, R., {de Bruijne}, J.~H.~J., {Brown}, A.~G.~A.,
  \& {Blaauw}, A. 1999, \aj, 117, 354, \dodoi{10.1086/300682}

\bibitem[{{Dhital} {et~al.}(2011){Dhital}, {Burgasser}, {Looper}, \&
  {Stassun}}]{dhital2011}
{Dhital}, S., {Burgasser}, A.~J., {Looper}, D.~L., \& {Stassun}, K.~G. 2011,
  \aj, 141, 7, \dodoi{10.1088/0004-6256/141/1/7}

\bibitem[{{Dohlen} {et~al.}(2008){Dohlen}, {Langlois}, {Saisse}, {Hill},
  {Origne}, {Jacquet}, {Fabron}, {Blanc}, {Llored}, {Carle}, {Moutou}, {Vigan},
  {Boccaletti}, {Carbillet}, {Mouillet}, \& {Beuzit}}]{dohlen2008}
{Dohlen}, K., {Langlois}, M., {Saisse}, M., {et~al.} 2008, in \procspie, Vol.
  7014, Ground-based and Airborne Instrumentation for Astronomy II, 70143L,
  \dodoi{10.1117/12.789786}

\bibitem[{{Faherty} {et~al.}(2009){Faherty}, {Burgasser}, {Cruz}, {Shara},
  {Walter}, \& {Gelino}}]{faherty2009}
{Faherty}, J.~K., {Burgasser}, A.~J., {Cruz}, K.~L., {et~al.} 2009, \aj, 137,
  1, \dodoi{10.1088/0004-6256/137/1/1}

\bibitem[{{Foreman-Mackey} {et~al.}(2013){Foreman-Mackey}, {Hogg}, {Lang}, \&
  {Goodman}}]{foreman-mackey2013}
{Foreman-Mackey}, D., {Hogg}, D.~W., {Lang}, D., \& {Goodman}, J. 2013, \pasp,
  125, 306, \dodoi{10.1086/670067}

\bibitem[{{Fusco} {et~al.}(2006){Fusco}, {Rousset}, {Sauvage}, {Petit},
  {Beuzit}, {Dohlen}, {Mouillet}, {Charton}, {Nicolle}, {Kasper}, {Baudoz}, \&
  {Puget}}]{fusco2006}
{Fusco}, T., {Rousset}, G., {Sauvage}, J.-F., {et~al.} 2006, Optics Express,
  14, 7515, \dodoi{10.1364/OE.14.007515}

\bibitem[{{Gaia Collaboration} {et~al.}(2018){Gaia Collaboration}, {Brown},
  {Vallenari}, {Prusti}, {de Bruijne}, {Babusiaux}, {Bailer-Jones}, {Biermann},
  {Evans}, {Eyer}, {Jansen}, {Jordi}, {Klioner}, {Lammers}, {Lindegren},
  {Luri}, {Mignard}, {Panem}, {Pourbaix}, {Randich}, {Sartoretti}, {Siddiqui},
  {Soubiran}, {van Leeuwen}, {Walton}, {Arenou}, {Bastian}, {Cropper},
  {Drimmel}, {Katz}, {Lattanzi}, {Bakker}, {Cacciari}, {Casta{\~n}eda},
  {Chaoul}, {Cheek}, {De Angeli}, {Fabricius}, {Guerra}, {Holl}, {Masana},
  {Messineo}, {Mowlavi}, {Nienartowicz}, {Panuzzo}, {Portell}, {Riello},
  {Seabroke}, {Tanga}, {Th{\'e}venin}, {Gracia-Abril}, {Comoretto},
  {Garcia-Reinaldos}, {Teyssier}, {Altmann}, {Andrae}, {Audard},
  {Bellas-Velidis}, {Benson}, {Berthier}, {Blomme}, {Burgess}, {Busso},
  {Carry}, {Cellino}, {Clementini}, {Clotet}, {Creevey}, {Davidson}, {De
  Ridder}, {Delchambre}, {Dell'Oro}, {Ducourant},
  {Fern{\'a}ndez-Hern{\'a}ndez}, {Fouesneau}, {Fr{\'e}mat}, {Galluccio},
  {Garc{\'\i}a-Torres}, {Gonz{\'a}lez-N{\'u}{\~n}ez}, {Gonz{\'a}lez-Vidal},
  {Gosset}, {Guy}, {Halbwachs}, {Hambly}, {Harrison}, {Hern{\'a}ndez},
  {Hestroffer}, {Hodgkin}, {Hutton}, {Jasniewicz}, {Jean-Antoine-Piccolo},
  {Jordan}, {Korn}, {Krone-Martins}, {Lanzafame}, {Lebzelter}, {L{\"o}ffler},
  {Manteiga}, {Marrese}, {Mart{\'\i}n-Fleitas}, {Moitinho}, {Mora}, {Muinonen},
  {Osinde}, {Pancino}, {Pauwels}, {Petit}, {Recio-Blanco}, {Richards},
  {Rimoldini}, {Robin}, {Sarro}, {Siopis}, {Smith}, {Sozzetti}, {S{\"u}veges},
  {Torra}, {van Reeven}, {Abbas}, {Abreu Aramburu}, {Accart}, {Aerts},
  {Altavilla}, {{\'A}lvarez}, {Alvarez}, {Alves}, {Anderson}, {Andrei},
  {Anglada Varela}, {Antiche}, {Antoja}, {Arcay}, {Astraatmadja}, {Bach},
  {Baker}, {Balaguer-N{\'u}{\~n}ez}, {Balm}, {Barache}, {Barata}, {Barbato},
  {Barblan}, {Barklem}, {Barrado}, {Barros}, {Barstow}, {Bartholom{\'e}
  Mu{\~n}oz}, {Bassilana}, {Becciani}, {Bellazzini}, {Berihuete}, {Bertone},
  {Bianchi}, {Bienaym{\'e}}, {Blanco-Cuaresma}, {Boch}, {Boeche}, {Bombrun},
  {Borrachero}, {Bossini}, {Bouquillon}, {Bourda}, {Bragaglia}, {Bramante},
  {Breddels}, {Bressan}, {Brouillet}, {Br{\"u}semeister}, {Brugaletta},
  {Bucciarelli}, {Burlacu}, {Busonero}, {Butkevich}, {Buzzi}, {Caffau},
  {Cancelliere}, {Cannizzaro}, {Cantat-Gaudin}, {Carballo}, {Carlucci},
  {Carrasco}, {Casamiquela}, {Castellani}, {Castro-Ginard}, {Charlot},
  {Chemin}, {Chiavassa}, {Cocozza}, {Costigan}, {Cowell}, {Crifo}, {Crosta},
  {Crowley}, {Cuypers}, {Dafonte}, {Damerdji}, {Dapergolas}, {David}, {David},
  {de Laverny}, {De Luise}, {De March}, {de Martino}, {de Souza}, {de Torres},
  {Debosscher}, {del Pozo}, {Delbo}, {Delgado}, {Delgado}, {Di Matteo},
  {Diakite}, {Diener}, {Distefano}, {Dolding}, {Drazinos}, {Dur{\'a}n},
  {Edvardsson}, {Enke}, {Eriksson}, {Esquej}, {Eynard Bontemps}, {Fabre},
  {Fabrizio}, {Faigler}, {Falc{\~a}o}, {Farr{\`a}s Casas}, {Federici},
  {Fedorets}, {Fernique}, {Figueras}, {Filippi}, {Findeisen}, {Fonti},
  {Fraile}, {Fraser}, {Fr{\'e}zouls}, {Gai}, {Galleti}, {Garabato},
  {Garc{\'\i}a-Sedano}, {Garofalo}, {Garralda}, {Gavel}, {Gavras}, {Gerssen},
  {Geyer}, {Giacobbe}, {Gilmore}, {Girona}, {Giuffrida}, {Glass}, {Gomes},
  {Granvik}, {Gueguen}, {Guerrier}, {Guiraud}, {Guti{\'e}rrez-S{\'a}nchez},
  {Haigron}, {Hatzidimitriou}, {Hauser}, {Haywood}, {Heiter}, {Helmi}, {Heu},
  {Hilger}, {Hobbs}, {Hofmann}, {Holland}, {Huckle}, {Hypki}, {Icardi},
  {Jan{\ss}en}, {Jevardat de Fombelle}, {Jonker}, {Juh{\'a}sz}, {Julbe},
  {Karampelas}, {Kewley}, {Klar}, {Kochoska}, {Kohley}, {Kolenberg},
  {Kontizas}, {Kontizas}, {Koposov}, {Kordopatis}, {Kostrzewa-Rutkowska},
  {Koubsky}, {Lambert}, {Lanza}, {Lasne}, {Lavigne}, {Le Fustec}, {Le
  Poncin-Lafitte}, {Lebreton}, {Leccia}, {Leclerc}, {Lecoeur-Taibi},
  {Lenhardt}, {Leroux}, {Liao}, {Licata}, {Lindstr{\o}m}, {Lister}, {Livanou},
  {Lobel}, {L{\'o}pez}, {Managau}, {Mann}, {Mantelet}, {Marchal}, {Marchant},
  {Marconi}, {Marinoni}, {Marschalk{\'o}}, {Marshall}, {Martino}, {Marton},
  {Mary}, {Massari}, {Matijevi{\v{c}}}, {Mazeh}, {McMillan}, {Messina},
  {Michalik}, {Millar}, {Molina}, {Molinaro}, {Moln{\'a}r}, {Montegriffo},
  {Mor}, {Morbidelli}, {Morel}, {Morris}, {Mulone}, {Muraveva}, {Musella},
  {Nelemans}, {Nicastro}, {Noval}, {O'Mullane}, {Ord{\'e}novic},
  {Ord{\'o}{\~n}ez-Blanco}, {Osborne}, {Pagani}, {Pagano}, {Pailler},
  {Palacin}, {Palaversa}, {Panahi}, {Pawlak}, {Piersimoni}, {Pineau}, {Plachy},
  {Plum}, {Poggio}, {Poujoulet}, {Pr{\v{s}}a}, {Pulone}, {Racero}, {Ragaini},
  {Rambaux}, {Ramos-Lerate}, {Regibo}, {Reyl{\'e}}, {Riclet}, {Ripepi}, {Riva},
  {Rivard}, {Rixon}, {Roegiers}, {Roelens}, {Romero-G{\'o}mez}, {Rowell},
  {Royer}, {Ruiz-Dern}, {Sadowski}, {Sagrist{\`a} Sell{\'e}s}, {Sahlmann},
  {Salgado}, {Salguero}, {Sanna}, {Santana-Ros}, {Sarasso}, {Savietto},
  {Schultheis}, {Sciacca}, {Segol}, {Segovia}, {S{\'e}gransan}, {Shih},
  {Siltala}, {Silva}, {Smart}, {Smith}, {Solano}, {Solitro}, {Sordo}, {Soria
  Nieto}, {Souchay}, {Spagna}, {Spoto}, {Stampa}, {Steele},
  {Steidelm{\"u}ller}, {Stephenson}, {Stoev}, {Suess}, {Surdej}, {Szabados},
  {Szegedi-Elek}, {Tapiador}, {Taris}, {Tauran}, {Taylor}, {Teixeira},
  {Terrett}, {Teyssand ier}, {Thuillot}, {Titarenko}, {Torra Clotet}, {Turon},
  {Ulla}, {Utrilla}, {Uzzi}, {Vaillant}, {Valentini}, {Valette}, {van Elteren},
  {Van Hemelryck}, {van Leeuwen}, {Vaschetto}, {Vecchiato}, {Veljanoski},
  {Viala}, {Vicente}, {Vogt}, {von Essen}, {Voss}, {Votruba}, {Voutsinas},
  {Walmsley}, {Weiler}, {Wertz}, {Wevers}, {Wyrzykowski}, {Yoldas},
  {{\v{Z}}erjal}, {Ziaeepour}, {Zorec}, {Zschocke}, {Zucker}, {Zurbach}, \&
  {Zwitter}}]{gaia2018}
{Gaia Collaboration}, {Brown}, A.~G.~A., {Vallenari}, A., {et~al.} 2018, \aap,
  616, A1, \dodoi{10.1051/0004-6361/201833051}

\bibitem[{{Galicher} {et~al.}(2014){Galicher}, {Rameau}, {Bonnefoy}, {Baudino},
  {Currie}, {Boccaletti}, {Chauvin}, {Lagrange}, \& {Marois}}]{galicher2014}
{Galicher}, R., {Rameau}, J., {Bonnefoy}, M., {et~al.} 2014, \aap, 565, L4,
  \dodoi{10.1051/0004-6361/201423839}

\bibitem[{{Gelino} \& {Burgasser}(2010)}]{gelino2010}
{Gelino}, C.~R., \& {Burgasser}, A.~J. 2010, \aj, 140, 110,
  \dodoi{10.1088/0004-6256/140/1/110}

\bibitem[{{Golimowski} {et~al.}(2004){Golimowski}, {Leggett}, {Marley}, {Fan},
  {Geballe}, {Knapp}, {Vrba}, {Henden}, {Luginbuhl}, {Guetter}, {Munn},
  {Canzian}, {Zheng}, {Tsvetanov}, {Chiu}, {Glazebrook}, {Hoversten},
  {Schneider}, \& {Brinkmann}}]{golimowski2004}
{Golimowski}, D.~A., {Leggett}, S.~K., {Marley}, M.~S., {et~al.} 2004, \aj,
  127, 3516, \dodoi{10.1086/420709}

\bibitem[{{Haffert} {et~al.}(2019){Haffert}, {Bohn}, {de Boer}, {Snellen},
  {Brinchmann}, {Girard}, {Keller}, \& {Bacon}}]{haffert2019}
{Haffert}, S.~Y., {Bohn}, A.~J., {de Boer}, J., {et~al.} 2019, Nature
  Astronomy, 3, 749, \dodoi{10.1038/s41550-019-0780-5}

\bibitem[{{Hom} {et~al.}(2020){Hom}, {Patience}, {Esposito}, {Duch{\^e}ne},
  {Worthen}, {Kalas}, {Jang-Condell}, {Saboi}, {Arriaga}, {Mazoyer}, {Wolff},
  {Millar-Blanchaer}, {Fitzgerald}, {Perrin}, {Chen}, {Macintosh}, {Matthews},
  {Wang}, {Graham}, {Marchis}, {Ammons}, {Bailey}, {Barman}, {Bulger},
  {Chilcote}, {Cotten}, {De Rosa}, {Doyon}, {Follette}, {Goodsell},
  {Greenbaum}, {Hibon}, {Ingraham}, {Konopacky}, {Larkin}, {Maire}, {Marley},
  {Marois}, {Matthews}, {Metchev}, {Nielsen}, {Oppenheimer}, {Palmer},
  {Poyneer}, {Pueyo}, {Rajan}, {Rameau}, {Rantakyr{\"o}}, {Ren}, {Savransky},
  {Schneider}, {Sivaramakrishnan}, {Song}, {Soummer}, {Tallis}, {Thomas},
  {Wallace}, {Ward-Duong}, {Wiktorowicz}, \& {Zuckerman}}]{Hom2020}
{Hom}, J., {Patience}, J., {Esposito}, T.~M., {et~al.} 2020, \aj, 159, 31,
  \dodoi{10.3847/1538-3881/ab5af2}

\bibitem[{{Hunter}(2007)}]{Matplotlib}
{Hunter}, J.~D. 2007, Computing in Science and Engineering, 9, 90,
  \dodoi{10.1109/MCSE.2007.55}

\bibitem[{{Janson} {et~al.}(2019){Janson}, {Asensio-Torres}, {Andr{\'e}},
  {Bonnefoy}, {Delorme}, {Reffert}, {Desidera}, {Langlois}, {Chauvin},
  {Gratton}, {Bohn}, {Eriksson}, {Marleau}, {Mamajek}, {Vigan}, \&
  {Carson}}]{janson2019}
{Janson}, M., {Asensio-Torres}, R., {Andr{\'e}}, D., {et~al.} 2019, \aap, 626,
  A99, \dodoi{10.1051/0004-6361/201935687}

\bibitem[{{Keppler} {et~al.}(2018){Keppler}, {Benisty}, {M{\"u}ller},
  {Henning}, {van Boekel}, {Cantalloube}, {Ginski}, {van Holstein}, {Maire},
  {Pohl}, {Samland}, {Avenhaus}, {Baudino}, {Boccaletti}, {de Boer},
  {Bonnefoy}, {Chauvin}, {Desidera}, {Langlois}, {Lazzoni}, {Marleau},
  {Mordasini}, {Pawellek}, {Stolker}, {Vigan}, {Zurlo}, {Birnstiel},
  {Brandner}, {Feldt}, {Flock}, {Girard}, {Gratton}, {Hagelberg}, {Isella},
  {Janson}, {Juhasz}, {Kemmer}, {Kral}, {Lagrange}, {Launhardt}, {Matter},
  {M{\'e}nard}, {Milli}, {Molli{\`e}re}, {Olofsson}, {P{\'e}rez}, {Pinilla},
  {Pinte}, {Quanz}, {Schmidt}, {Udry}, {Wahhaj}, {Williams}, {Buenzli},
  {Cudel}, {Dominik}, {Galicher}, {Kasper}, {Lannier}, {Mesa}, {Mouillet},
  {Peretti}, {Perrot}, {Salter}, {Sissa}, {Wildi}, {Abe}, {Antichi},
  {Augereau}, {Baruffolo}, {Baudoz}, {Bazzon}, {Beuzit}, {Blanchard}, {Brems},
  {Buey}, {De Caprio}, {Carbillet}, {Carle}, {Cascone}, {Cheetham}, {Claudi},
  {Costille}, {Delboulb{\'e}}, {Dohlen}, {Fantinel}, {Feautrier}, {Fusco},
  {Giro}, {Gluck}, {Gry}, {Hubin}, {Hugot}, {Jaquet}, {Le Mignant}, {Llored},
  {Madec}, {Magnard}, {Martinez}, {Maurel}, {Meyer}, {M{\"o}ller-Nilsson},
  {Moulin}, {Mugnier}, {Orign{\'e}}, {Pavlov}, {Perret}, {Petit}, {Pragt},
  {Puget}, {Rabou}, {Ramos}, {Rigal}, {Rochat}, {Roelfsema}, {Rousset}, {Roux},
  {Salasnich}, {Sauvage}, {Sevin}, {Soenke}, {Stadler}, {Suarez}, {Turatto}, \&
  {Weber}}]{keppler2018}
{Keppler}, M., {Benisty}, M., {M{\"u}ller}, A., {et~al.} 2018, \aap, 617, A44,
  \dodoi{10.1051/0004-6361/201832957}

\bibitem[{{Kirkpatrick} {et~al.}(2006){Kirkpatrick}, {Barman}, {Burgasser},
  {McGovern}, {McLean}, {Tinney}, \& {Lowrance}}]{kirkpatrick2006}
{Kirkpatrick}, J.~D., {Barman}, T.~S., {Burgasser}, A.~J., {et~al.} 2006, \apj,
  639, 1120, \dodoi{10.1086/499622}

\bibitem[{{Kirkpatrick} {et~al.}(2010){Kirkpatrick}, {Looper}, {Burgasser},
  {Schurr}, {Cutri}, {Cushing}, {Cruz}, {Sweet}, {Knapp}, {Barman},
  {Bochanski}, {Roellig}, {McLean}, {McGovern}, \& {Rice}}]{kirkpatrick2010}
{Kirkpatrick}, J.~D., {Looper}, D.~L., {Burgasser}, A.~J., {et~al.} 2010,
  \apjs, 190, 100, \dodoi{10.1088/0067-0049/190/1/100}

\bibitem[{{Knapp} {et~al.}(2004){Knapp}, {Leggett}, {Fan}, {Marley}, {Geballe},
  {Golimowski}, {Finkbeiner}, {Gunn}, {Hennawi}, {Ivezi{\'c}}, {Lupton},
  {Schlegel}, {Strauss}, {Tsvetanov}, {Chiu}, {Hoversten}, {Glazebrook},
  {Zheng}, {Hendrickson}, {Williams}, {Uomoto}, {Vrba}, {Henden}, {Luginbuhl},
  {Guetter}, {Munn}, {Canzian}, {Schneider}, \& {Brinkmann}}]{knapp2004}
{Knapp}, G.~R., {Leggett}, S.~K., {Fan}, X., {et~al.} 2004, \aj, 127, 3553,
  \dodoi{10.1086/420707}

\bibitem[{{Lafreni{\`e}re} {et~al.}(2008){Lafreni{\`e}re}, {Jayawardhana}, \&
  {van Kerkwijk}}]{lafreniere2008}
{Lafreni{\`e}re}, D., {Jayawardhana}, R., \& {van Kerkwijk}, M.~H. 2008, The
  Astrophysical Journal, 689, L153, \dodoi{10.1086/595870}

\bibitem[{{Lenzen} {et~al.}(2003){Lenzen}, {Hartung}, {Brandner}, {Finger},
  {Hubin}, {Lacombe}, {Lagrange}, {Lehnert}, {Moorwood}, \&
  {Mouillet}}]{lenzen2003}
{Lenzen}, R., {Hartung}, M., {Brandner}, W., {et~al.} 2003, in Society of
  Photo-Optical Instrumentation Engineers (SPIE) Conference Series, Vol. 4841,
  Instrument Design and Performance for Optical/Infrared Ground-based
  Telescopes, ed. M.~{Iye} \& A.~F.~M. {Moorwood}, 944--952,
  \dodoi{10.1117/12.460044}

\bibitem[{{Looper} {et~al.}(2010){Looper}, {Bochanski}, {Burgasser}, {Mohanty},
  {Mamajek}, {Faherty}, {West}, \& {Pitts}}]{looper2010}
{Looper}, D.~L., {Bochanski}, J.~J., {Burgasser}, A.~J., {et~al.} 2010, \aj,
  140, 1486, \dodoi{10.1088/0004-6256/140/5/1486}

\bibitem[{{Looper} {et~al.}(2007){Looper}, {Burgasser}, {Kirkpatrick}, \&
  {Swift}}]{looper2007}
{Looper}, D.~L., {Burgasser}, A.~J., {Kirkpatrick}, J.~D., \& {Swift}, B.~J.
  2007, \apjl, 669, L97, \dodoi{10.1086/523812}

\bibitem[{{Macintosh} {et~al.}(2014){Macintosh}, {Graham}, {Ingraham},
  {Konopacky}, {Marois}, {Perrin}, {Poyneer}, {Bauman}, {Barman}, {Burrows},
  {Cardwell}, {Chilcote}, {De Rosa}, {Dillon}, {Doyon}, {Dunn}, {Erikson},
  {Fitzgerald}, {Gavel}, {Goodsell}, {Hartung}, {Hibon}, {Kalas}, {Larkin},
  {Maire}, {Marchis}, {Marley}, {McBride}, {Millar-Blanchaer}, {Morzinski},
  {Norton}, {Oppenheimer}, {Palmer}, {Patience}, {Pueyo}, {Rantakyro},
  {Sadakuni}, {Saddlemyer}, {Savransky}, {Serio}, {Soummer},
  {Sivaramakrishnan}, {Song}, {Thomas}, {Wallace}, {Wiktorowicz}, \&
  {Wolff}}]{macintosh2014}
{Macintosh}, B., {Graham}, J.~R., {Ingraham}, P., {et~al.} 2014, Proceedings of
  the National Academy of Science, 111, 12661, \dodoi{10.1073/pnas.1304215111}

\bibitem[{{Macintosh} {et~al.}(2015){Macintosh}, {Graham}, {Barman}, {De Rosa},
  {Konopacky}, {Marley}, {Marois}, {Nielsen}, {Pueyo}, {Rajan}, {Rameau},
  {Saumon}, {Wang}, {Patience}, {Ammons}, {Arriaga}, {Artigau}, {Beckwith},
  {Brewster}, {Bruzzone}, {Bulger}, {Burningham}, {Burrows}, {Chen}, {Chiang},
  {Chilcote}, {Dawson}, {Dong}, {Doyon}, {Draper}, {Duch{\^e}ne}, {Esposito},
  {Fabrycky}, {Fitzgerald}, {Follette}, {Fortney}, {Gerard}, {Goodsell},
  {Greenbaum}, {Hibon}, {Hinkley}, {Cotten}, {Hung}, {Ingraham},
  {Johnson-Groh}, {Kalas}, {Lafreniere}, {Larkin}, {Lee}, {Line}, {Long},
  {Maire}, {Marchis}, {Matthews}, {Max}, {Metchev}, {Millar-Blanchaer},
  {Mittal}, {Morley}, {Morzinski}, {Murray-Clay}, {Oppenheimer}, {Palmer},
  {Patel}, {Perrin}, {Poyneer}, {Rafikov}, {Rantakyr{\"o}}, {Rice}, {Rojo},
  {Rudy}, {Ruffio}, {Ruiz}, {Sadakuni}, {Saddlemyer}, {Salama}, {Savransky},
  {Schneider}, {Sivaramakrishnan}, {Song}, {Soummer}, {Thomas}, {Vasisht},
  {Wallace}, {Ward- Duong}, {Wiktorowicz}, {Wolff}, \&
  {Zuckerman}}]{macintosh2015}
{Macintosh}, B., {Graham}, J.~R., {Barman}, T., {et~al.} 2015, Science, 350,
  64, \dodoi{10.1126/science.aac5891}

\bibitem[{{Maire} {et~al.}(2016){Maire}, {Langlois}, {Dohlen}, {Lagrange},
  {Gratton}, {Chauvin}, {Desidera}, {Girard}, {Milli}, {Vigan}, {Zins},
  {Delorme}, {Beuzit}, {Claudi}, {Feldt}, {Mouillet}, {Puget}, {Turatto}, \&
  {Wildi}}]{maire2016}
{Maire}, A.-L., {Langlois}, M., {Dohlen}, K., {et~al.} 2016, in Society of
  Photo-Optical Instrumentation Engineers (SPIE) Conference Series, Vol. 9908,
  \procspie, 990834, \dodoi{10.1117/12.2233013}

\bibitem[{{Marois} {et~al.}(2008){Marois}, {Macintosh}, {Barman}, {Zuckerman},
  {Song}, {Patience}, {Lafreni{\`e}re}, \& {Doyon}}]{marois2008}
{Marois}, C., {Macintosh}, B., {Barman}, T., {et~al.} 2008, Science, 322, 1348,
  \dodoi{10.1126/science.1166585}

\bibitem[{{Marois} {et~al.}(2010){Marois}, {Zuckerman}, {Konopacky},
  {Macintosh}, \& {Barman}}]{marois2010}
{Marois}, C., {Zuckerman}, B., {Konopacky}, Q.~M., {Macintosh}, B., \&
  {Barman}, T. 2010, \nat, 468, 1080, \dodoi{10.1038/nature09684}

\bibitem[{{Mawet} {et~al.}(2014){Mawet}, {Milli}, {Wahhaj}, {Pelat}, {Absil},
  {Delacroix}, {Boccaletti}, {Kasper}, {Kenworthy}, {Marois}, {Mennesson}, \&
  {Pueyo}}]{mawet2014}
{Mawet}, D., {Milli}, J., {Wahhaj}, Z., {et~al.} 2014, \apj, 792, 97,
  \dodoi{10.1088/0004-637X/792/2/97}

\bibitem[{{McElwain} \& {Burgasser}(2006)}]{mcelwain2006}
{McElwain}, M.~W., \& {Burgasser}, A.~J. 2006, \aj, 132, 2074,
  \dodoi{10.1086/508199}

\bibitem[{{McLean} {et~al.}(2003){McLean}, {McGovern}, {Burgasser},
  {Kirkpatrick}, {Prato}, \& {Kim}}]{mclean2003}
{McLean}, I.~S., {McGovern}, M.~R., {Burgasser}, A.~J., {et~al.} 2003, \apj,
  596, 561, \dodoi{10.1086/377636}

\bibitem[{{McLean} {et~al.}(2007){McLean}, {Prato}, {McGovern}, {Burgasser},
  {Kirkpatrick}, {Rice}, \& {Kim}}]{mclean2007}
{McLean}, I.~S., {Prato}, L., {McGovern}, M.~R., {et~al.} 2007, \apj, 658,
  1217, \dodoi{10.1086/511740}

\bibitem[{{Mesa} {et~al.}(2019){Mesa}, {Keppler}, {Cantalloube}, {Rodet},
  {Charnay}, {Gratton}, {Langlois}, {Boccaletti}, {Bonnefoy}, {Vigan},
  {Flasseur}, {Bae}, {Benisty}, {Chauvin}, {de Boer}, {Desidera}, {Henning},
  {Lagrange}, {Meyer}, {Milli}, {M{\"u}ller}, {Pairet}, {Zurlo}, {Antoniucci},
  {Baudino}, {Brown Sevilla}, {Cascone}, {Cheetham}, {Claudi}, {Delorme},
  {D'Orazi}, {Feldt}, {Hagelberg}, {Janson}, {Kral}, {Lagadec}, {Lazzoni},
  {Ligi}, {Maire}, {Martinez}, {Menard}, {Meunier}, {Perrot}, {Petrus},
  {Pinte}, {Rickman}, {Rochat}, {Rouan}, {Samland}, {Sauvage}, {Schmidt},
  {Udry}, {Weber}, \& {Wildi}}]{mesa2019}
{Mesa}, D., {Keppler}, M., {Cantalloube}, F., {et~al.} 2019, \aap, 632, A25,
  \dodoi{10.1051/0004-6361/201936764}

\bibitem[{{Mo{\'o}r} {et~al.}(2017){Mo{\'o}r}, {Cur{\'e}}, {K{\'o}sp{\'a}l},
  {{\'A}brah{\'a}m}, {Csengeri}, {Eiroa}, {Gunawan}, {Henning}, {Hughes},
  {Juh{\'a}sz}, {Pawellek}, \& {Wyatt}}]{moor2017}
{Mo{\'o}r}, A., {Cur{\'e}}, M., {K{\'o}sp{\'a}l}, {\'A}., {et~al.} 2017, \apj,
  849, 123, \dodoi{10.3847/1538-4357/aa8e4e}

\bibitem[{{Morbidelli}(2018)}]{morbidelli2018}
{Morbidelli}, A. 2018, {Dynamical Evolution of Planetary Systems}, 145,
  \dodoi{10.1007/978-3-319-55333-7_145}

\bibitem[{{Mordasini} {et~al.}(2012){Mordasini}, {Alibert}, {Georgy},
  {Dittkrist}, {Klahr}, \& {Henning}}]{mordasini2012}
{Mordasini}, C., {Alibert}, Y., {Georgy}, C., {et~al.} 2012, \aap, 547, A112,
  \dodoi{10.1051/0004-6361/201118464}

\bibitem[{{Muench} {et~al.}(2007){Muench}, {Lada}, {Luhman}, {Muzerolle}, \&
  {Young}}]{muench2007}
{Muench}, A.~A., {Lada}, C.~J., {Luhman}, K.~L., {Muzerolle}, J., \& {Young},
  E. 2007, \aj, 134, 411, \dodoi{10.1086/518560}

\bibitem[{{M{\"u}ller} {et~al.}(2018){M{\"u}ller}, {Keppler}, {Henning},
  {Samland}, {Chauvin}, {Beust}, {Maire}, {Molaverdikhani}, {van Boekel},
  {Benisty}, {Boccaletti}, {Bonnefoy}, {Cantalloube}, {Charnay}, {Baudino},
  {Gennaro}, {Long}, {Cheetham}, {Desidera}, {Feldt}, {Fusco}, {Girard},
  {Gratton}, {Hagelberg}, {Janson}, {Lagrange}, {Langlois}, {Lazzoni}, {Ligi},
  {M{\'e}nard}, {Mesa}, {Meyer}, {Molli{\`e}re}, {Mordasini}, {Moulin},
  {Pavlov}, {Pawellek}, {Quanz}, {Ramos}, {Rouan}, {Sissa}, {Stadler}, {Vigan},
  {Wahhaj}, {Weber}, \& {Zurlo}}]{muller2018}
{M{\"u}ller}, A., {Keppler}, M., {Henning}, T., {et~al.} 2018, \aap, 617, L2,
  \dodoi{10.1051/0004-6361/201833584}

\bibitem[{Nelder \& Mead(1965)}]{nelder1965}
Nelder, J.~A., \& Mead, R. 1965, The computer journal, 7, 308

\bibitem[{Oliphant(2006)}]{numpy}
Oliphant, T.~E. 2006, A guide to NumPy, Vol.~1 (Trelgol Publishing USA)

\bibitem[{{Pecaut} \& {Mamajek}(2016)}]{pecaut2016}
{Pecaut}, M.~J., \& {Mamajek}, E.~E. 2016, \mnras, 461, 794,
  \dodoi{10.1093/mnras/stw1300}

\bibitem[{{Pedregosa} {et~al.}(2012){Pedregosa}, {Varoquaux}, {Gramfort},
  {Michel}, {Thirion}, {Grisel}, {Blondel}, {M{\"u}ller}, {Nothman}, {Louppe},
  {Prettenhofer}, {Weiss}, {Dubourg}, {Vanderplas}, {Passos}, {Cournapeau},
  {Brucher}, {Perrot}, \& {Duchesnay}}]{scikit-learn}
{Pedregosa}, F., {Varoquaux}, G., {Gramfort}, A., {et~al.} 2012, arXiv
  e-prints, arXiv:1201.0490.
\newblock \doarXiv{1201.0490}

\bibitem[{{Pelupessy} {et~al.}(2013){Pelupessy}, {van Elteren}, {de Vries},
  {McMillan}, {Drost}, \& {Portegies Zwart}}]{amuse}
{Pelupessy}, F.~I., {van Elteren}, A., {de Vries}, N., {et~al.} 2013, \aap,
  557, A84, \dodoi{10.1051/0004-6361/201321252}

\bibitem[{{Rayner} {et~al.}(2009){Rayner}, {Cushing}, \& {Vacca}}]{rayner2009}
{Rayner}, J.~T., {Cushing}, M.~C., \& {Vacca}, W.~D. 2009, \apjs, 185, 289,
  \dodoi{10.1088/0067-0049/185/2/289}

\bibitem[{{Reid} {et~al.}(2006){Reid}, {Lewitus}, {Burgasser}, \&
  {Cruz}}]{reid2006}
{Reid}, I.~N., {Lewitus}, E., {Burgasser}, A.~J., \& {Cruz}, K.~L. 2006, \apj,
  639, 1114, \dodoi{10.1086/499484}

\bibitem[{{Rein} \& {Liu}(2012)}]{rebound}
{Rein}, H., \& {Liu}, S.~F. 2012, \aap, 537, A128,
  \dodoi{10.1051/0004-6361/201118085}

\bibitem[{{Rein} \& {Tamayo}(2015)}]{whfast}
{Rein}, H., \& {Tamayo}, D. 2015, \mnras, 452, 376,
  \dodoi{10.1093/mnras/stv1257}

\bibitem[{{Rein} \& {Tamayo}(2016)}]{megno}
---. 2016, \mnras, 459, 2275, \dodoi{10.1093/mnras/stw644}

\bibitem[{{Rousset} {et~al.}(2003){Rousset}, {Lacombe}, {Puget}, {Hubin},
  {Gendron}, {Fusco}, {Arsenault}, {Charton}, {Feautrier}, \&
  {Gigan}}]{rousset2003}
{Rousset}, G., {Lacombe}, F., {Puget}, P., {et~al.} 2003, in Society of
  Photo-Optical Instrumentation Engineers (SPIE) Conference Series, Vol. 4839,
  Adaptive Optical System Technologies II, ed. P.~L. {Wizinowich} \&
  D.~{Bonaccini}, 140--149, \dodoi{10.1117/12.459332}

\bibitem[{{Samland} {et~al.}(2017){Samland}, {Molli{\`e}re}, {Bonnefoy},
  {Maire}, {Cantalloube}, {Cheetham}, {Mesa}, {Gratton}, {Biller}, {Wahhaj},
  {Bouwman}, {Brandner}, {Melnick}, {Carson}, {Janson}, {Henning}, {Homeier},
  {Mordasini}, {Langlois}, {Quanz}, {van Boekel}, {Zurlo}, {Schlieder},
  {Avenhaus}, {Beuzit}, {Boccaletti}, {Bonavita}, {Chauvin}, {Claudi}, {Cudel},
  {Desidera}, {Feldt}, {Fusco}, {Galicher}, {Kopytova}, {Lagrange}, {Le
  Coroller}, {Martinez}, {Moeller-Nilsson}, {Mouillet}, {Mugnier}, {Perrot},
  {Sevin}, {Sissa}, {Vigan}, \& {Weber}}]{samland2017}
{Samland}, M., {Molli{\`e}re}, P., {Bonnefoy}, M., {et~al.} 2017, \aap, 603,
  A57, \dodoi{10.1051/0004-6361/201629767}

\bibitem[{{Sheppard} \& {Cushing}(2009)}]{sheppard2009}
{Sheppard}, S.~S., \& {Cushing}, M.~C. 2009, \aj, 137, 304,
  \dodoi{10.1088/0004-6256/137/1/304}

\bibitem[{{Siegler} {et~al.}(2007){Siegler}, {Close}, {Burgasser}, {Cruz},
  {Marois}, {Macintosh}, \& {Barman}}]{siegler2007}
{Siegler}, N., {Close}, L.~M., {Burgasser}, A.~J., {et~al.} 2007, \aj, 133,
  2320, \dodoi{10.1086/513273}

\bibitem[{{Snellen} {et~al.}(2014){Snellen}, {Brandl}, {de Kok}, {Brogi},
  {Birkby}, \& {Schwarz}}]{snellen2014}
{Snellen}, I. A.~G., {Brandl}, B.~R., {de Kok}, R.~J., {et~al.} 2014, \nat,
  509, 63, \dodoi{10.1038/nature13253}

\bibitem[{{Soummer} {et~al.}(2012){Soummer}, {Pueyo}, \&
  {Larkin}}]{soummer2012}
{Soummer}, R., {Pueyo}, L., \& {Larkin}, J. 2012, \apj, 755, L28,
  \dodoi{10.1088/2041-8205/755/2/L28}

\bibitem[{{Stolker} {et~al.}(2019){Stolker}, {Bonse}, {Quanz}, {Amara},
  {Cugno}, {Bohn}, \& {Boehle}}]{stolker2019}
{Stolker}, T., {Bonse}, M.~J., {Quanz}, S.~P., {et~al.} 2019, \aap, 621, A59,
  \dodoi{10.1051/0004-6361/201834136}

\bibitem[{Van~der Walt {et~al.}(2014)Van~der Walt, Sch{\"o}nberger,
  Nunez-Iglesias, Boulogne, Warner, Yager, Gouillart, \& Yu}]{scikit-image}
Van~der Walt, S., Sch{\"o}nberger, J.~L., Nunez-Iglesias, J., {et~al.} 2014,
  PeerJ, 2, e453

\bibitem[{{van Holstein} {et~al.}(2020){van Holstein}, {Girard}, {de Boer},
  {Snik}, {Milli}, {Stam}, {Ginski}, {Mouillet}, {Wahhaj}, {Schmid}, {Keller},
  {Langlois}, {Dohlen}, {Vigan}, {Pohl}, {Carbillet}, {Fantinel}, {Maurel},
  {Orign{\'e}}, {Petit}, {Ramos}, {Rigal}, {Sevin}, {Boccaletti}, {Le
  Coroller}, {Dominik}, {Henning}, {Lagadec}, {M{\'e}nard}, {Turatto}, {Udry},
  {Chauvin}, {Feldt}, \& {Beuzit}}]{vanHolstein2020}
{van Holstein}, R.~G., {Girard}, J.~H., {de Boer}, J., {et~al.} 2020, \aap,
  633, A64, \dodoi{10.1051/0004-6361/201834996}

\bibitem[{{Vigan} {et~al.}(2010){Vigan}, {Moutou}, {Langlois}, {Allard},
  {Boccaletti}, {Carbillet}, {Mouillet}, \& {Smith}}]{vigan2010}
{Vigan}, A., {Moutou}, C., {Langlois}, M., {et~al.} 2010, \mnras, 407, 71,
  \dodoi{10.1111/j.1365-2966.2010.16916.x}

\bibitem[{{Virtanen} {et~al.}(2020){Virtanen}, {Gommers}, {Oliphant},
  {Haberland}, {Reddy}, {Cournapeau}, {Burovski}, {Peterson}, {Weckesser},
  {Bright}, {van der Walt}, {Brett}, {Wilson}, {Millman}, {Mayorov}, {Nelson},
  {Jones}, {Kern}, {Larson}, {Carey}, {Polat}, {Feng}, {Moore}, {Vand erPlas},
  {Laxalde}, {Perktold}, {Cimrman}, {Henriksen}, {Quintero}, {Harris},
  {Archibald}, {Ribeiro}, {Pedregosa}, {van Mulbregt}, \& {SciPy 1. 0
  Contributors}}]{virtanen2020}
{Virtanen}, P., {Gommers}, R., {Oliphant}, T.~E., {et~al.} 2020, Nature
  Methods, 17, 261, \dodoi{10.1038/s41592-019-0686-2}

\bibitem[{{Wang} {et~al.}(2018){Wang}, {Graham}, {Dawson}, {Fabrycky}, {De
  Rosa}, {Pueyo}, {Konopacky}, {Macintosh}, {Marois}, {Chiang}, {Ammons},
  {Arriaga}, {Bailey}, {Barman}, {Bulger}, {Chilcote}, {Cotten}, {Doyon},
  {Duch{\^e}ne}, {Esposito}, {Fitzgerald}, {Follette}, {Gerard}, {Goodsell},
  {Greenbaum}, {Hibon}, {Hung}, {Ingraham}, {Kalas}, {Larkin}, {Maire},
  {Marchis}, {Marley}, {Metchev}, {Millar-Blanchaer}, {Nielsen}, {Oppenheimer},
  {Palmer}, {Patience}, {Perrin}, {Poyneer}, {Rajan}, {Rameau},
  {Rantakyr{\"o}}, {Ruffio}, {Savransky}, {Schneider}, {Sivaramakrishnan},
  {Song}, {Soummer}, {Thomas}, {Wallace}, {Ward-Duong}, {Wiktorowicz}, \&
  {Wolff}}]{wang2018}
{Wang}, J.~J., {Graham}, J.~R., {Dawson}, R., {et~al.} 2018, \aj, 156, 192,
  \dodoi{10.3847/1538-3881/aae150}

\bibitem[{{Wenger} {et~al.}(2000){Wenger}, {Ochsenbein}, {Egret}, {Dubois},
  {Bonnarel}, {Borde}, {Genova}, {Jasniewicz}, {Lalo{\"e}}, {Lesteven}, \&
  {Monier}}]{Wenger2000}
{Wenger}, M., {Ochsenbein}, F., {Egret}, D., {et~al.} 2000, \aaps, 143, 9,
  \dodoi{10.1051/aas:2000332}

\bibitem[{{Wertz} {et~al.}(2017){Wertz}, {Absil}, {G{\'o}mez Gonz{\'a}lez},
  {Milli}, {Girard}, {Mawet}, \& {Pueyo}}]{wertz2017}
{Wertz}, O., {Absil}, O., {G{\'o}mez Gonz{\'a}lez}, C.~A., {et~al.} 2017, \aap,
  598, A83, \dodoi{10.1051/0004-6361/201628730}

\bibitem[{{Zhou} {et~al.}(2016){Zhou}, {Apai}, {Schneider}, {Marley}, \&
  {Showman}}]{zhou2016}
{Zhou}, Y., {Apai}, D., {Schneider}, G.~H., {Marley}, M.~S., \& {Showman},
  A.~P. 2016, \apj, 818, 176, \dodoi{10.3847/0004-637X/818/2/176}

\bibitem[{{Zurlo} {et~al.}(2016){Zurlo}, {Vigan}, {Galicher}, {Maire}, {Mesa},
  {Gratton}, {Chauvin}, {Kasper}, {Moutou}, {Bonnefoy}, {Desidera}, {Abe},
  {Apai}, {Baruffolo}, {Baudoz}, {Baudrand}, {Beuzit}, {Blancard},
  {Boccaletti}, {Cantalloube}, {Carle}, {Cascone}, {Charton}, {Claudi},
  {Costille}, {de Caprio}, {Dohlen}, {Dominik}, {Fantinel}, {Feautrier},
  {Feldt}, {Fusco}, {Gigan}, {Girard}, {Gisler}, {Gluck}, {Gry}, {Henning},
  {Hugot}, {Janson}, {Jaquet}, {Lagrange}, {Langlois}, {Llored}, {Madec},
  {Magnard}, {Martinez}, {Maurel}, {Mawet}, {Meyer}, {Milli},
  {Moeller-Nilsson}, {Mouillet}, {Orign{\'e}}, {Pavlov}, {Petit}, {Puget},
  {Quanz}, {Rabou}, {Ramos}, {Rousset}, {Roux}, {Salasnich}, {Salter},
  {Sauvage}, {Schmid}, {Soenke}, {Stadler}, {Suarez}, {Turatto}, {Udry},
  {Vakili}, {Wahhaj}, {Wildi}, \& {Antichi}}]{zurlo2016}
{Zurlo}, A., {Vigan}, A., {Galicher}, R., {et~al.} 2016, \aap, 587, A57,
  \dodoi{10.1051/0004-6361/201526835}

\end{thebibliography}
\bibliographystyle{aasjournal}



\end{document}